\documentclass[twocolumn,aps,pra,showpacs,superscriptaddress,groupedaddress]{revtex4-1}
\usepackage[latin9]{inputenc}
\usepackage{amsmath}
\usepackage{amssymb}
\usepackage{graphicx}
\usepackage{color}
\usepackage{hyperref}
\graphicspath{ {main_figures/} }
\usepackage{color}
\usepackage{diagbox}
\usepackage{gensymb}
\usepackage{physics}
\usepackage{textcomp}

\newcommand{\commentOut}[1]{}

\makeatletter

\@ifundefined{textcolor}{}
{%
 \definecolor{BLACK}{gray}{0}
 \definecolor{WHITE}{gray}{1}
 \definecolor{RED}{rgb}{1,0,0}
 \definecolor{GREEN}{rgb}{0,1,0}
 \definecolor{BLUE}{rgb}{0,0,1}
 \definecolor{CYAN}{cmyk}{1,0,0,0}
 \definecolor{MAGENTA}{cmyk}{0,1,0,0}
 \definecolor{YELLOW}{cmyk}{0,0,1,0}
}

\usepackage{tabularx,booktabs}
\newcolumntype{Y}{>{\centering\arraybackslash}X}

\usepackage{dcolumn}
\usepackage{bm}
\makeatother

\begin{document}

\title{A Dissipatively Stabilized Mott Insulator of Photons}

\author{Ruichao Ma}
\address{James Franck Institute and Department of Physics, University of Chicago,
Chicago, Illinois 60637}
\author{Brendan Saxberg}
\address{James Franck Institute and Department of Physics, University of Chicago,
Chicago, Illinois 60637}
\author{Clai Owens}
\address{James Franck Institute and Department of Physics, University of Chicago,
Chicago, Illinois 60637}
\author{Nelson Leung}
\address{James Franck Institute and Department of Physics, University of Chicago,
Chicago, Illinois 60637}
\author{Yao Lu}
\address{James Franck Institute and Department of Physics, University of Chicago,
Chicago, Illinois 60637}
\author{Jonathan Simon}
\address{James Franck Institute and Department of Physics, University of Chicago,
Chicago, Illinois 60637}
\author{David I. Schuster}
\address{James Franck Institute and Department of Physics, University of Chicago,
Chicago, Illinois 60637}

\date{\today }

\begin{abstract}

Superconducting circuits are a competitive platform for quantum computation because they offer controllability, long coherence times and strong interactions---properties that are essential for the study of quantum materials comprising microwave photons. However, intrinsic photon losses in these circuits hinder the realization of quantum many-body phases. Here we use superconducting circuits to explore strongly correlated quantum matter by building a Bose-Hubbard lattice for photons in the strongly interacting regime. We develop a versatile method for dissipative preparation of incompressible many-body phases through reservoir engineering and apply it to our system to stabilize a Mott insulator of photons against losses. Site- and time-resolved readout of the lattice allows us to investigate the microscopic details of the thermalization process through the dynamics of defect propagation and removal in the Mott phase. Our experiments demonstrate the power of superconducting circuits for studying strongly correlated matter in both coherent and engineered dissipative settings. In conjunction with recently demonstrated superconducting microwave Chern insulators, we expect that our approach will enable the exploration of topologically ordered phases of matter.

\end{abstract}
\maketitle

The richness of quantum materials originates from the competition between quantum fluctuations arising from strong interactions, motional dynamics, and the topology of the system. The results of this competition manifest as strong correlations and entanglement, observed both in the equilibrium ground state and in non-equilibrium dynamical evolution. In most condensed matter systems, efficient thermalization with a cold environment and a well-defined chemical potential lead naturally to the preparation of the system near its many-body ground state, so understanding of the path to strong-correlations-- how particles order themselves under the system Hamiltonian-- often escapes notice.

Synthetic quantum materials provide an opportunity to investigate this paradigm. Built from highly coherent constituents with precisely controlled and tunable interactions and dynamics, they have emerged as ideal platforms to explore quantum correlations, owing to their slowed dynamics and capabilities in high-resolution imaging~\cite{Bakr2009,Sherson2010}. Low-entropy strongly-correlated states are typically reached adiabatically in a many-body analog of the Landau-Zener process by slowly tuning the system Hamiltonian through a quantum phase transition while isolated from the environment, starting with a low-entropy state prepared in a weakly interacting and/or weakly correlated regime. As a prominent example from atomic physics, laser- and evaporative- cooling remove entropy from weakly interacting atomic gases to create Bose Einstein condensates~\cite{anderson1995observation,davis1995bose} which are then used to adiabatically reach phases including Mott insulators~\cite{greiner2002quantum}, quantum magnets~\cite{Simon2010,mazurenko2016experimental}, and potentially even topologically ordered states~\cite{he2017realizing}. These coherent isolated systems have prompted exciting studies of relaxation in closed quantum systems, including observation of pre-thermalization~\cite{gring2012relaxation}, many-body localization~\cite{schreiber2015observation}, and quantum self-thermalization~\cite{kaufman2016quantum}. Nonetheless, the challenge in such a ``cool then adiabatically evolve'' approach is the competition between the limited coherence time and the adiabatic criterion at the smallest many-body gaps, which shrink in the quantum critical region and often vanish at topological phase transitions. This suggests dissipative stabilization of many-body states, which works \emph{directly} in the strongly-correlated phase with a potentially larger many-body gap, as a promising alternative approach. To date, though, thermalization into strongly correlated phases of synthetic matter has remained largely unexplored.

\begin{figure}[htbp]
    \includegraphics[width=0.95\columnwidth]{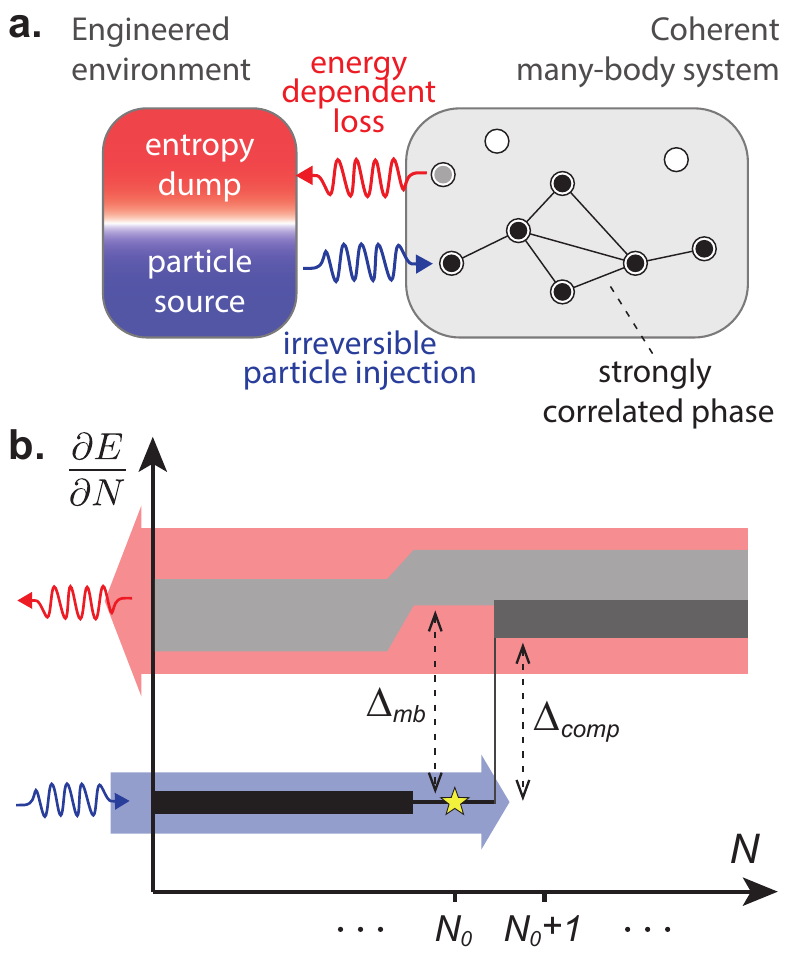}
    \caption{\textbf{Dissipative stabilization of incompressible many-body states.} \textbf{(a)} Illustration of using an engineered environment to populate a many-body system. \textbf{(b)} Photons are continuously added into the system irreversibly in a narrow band of energies (blue) that connect the initial vacuum to the desired target state (star). This process stops when the system is fully filled at photon number $N_0$, due to the presence of the compressibility gap $\Delta_{\textrm{comp}}$, thereby preparing and stabilizing the gapped (with $\Delta_{\textrm{mb}}$) many-body state where the photons can order into strongly correlated phases as a result of the underlying Hamiltonian. The energy dependent loss channels (red) ensure that any accidental excitations into higher states (gray) are short lived.
    }
    \label{fig:0}
\end{figure}

Recently, photonic systems have emerged as an exciting platform for exploration of synthetic quantum matter~\cite{Hartmann2006, Greentree2006, Angelakis2007, Noh2016, Hartmann2016, Gu2017}. In particular, superconducting circuits have been used to study many-body physics of microwave photons, leveraging the exquisite individual control of strongly interacting qubits. This approach builds upon the circuit quantum electrodynamics toolbox developed for quantum computing~\cite{wallraff2004strong}, and has been applied to digital simulation of spin models~\cite{salathe2015digital}, fermionic dynamics~\cite{barends2015digital} and quantum chemistry~\cite{Martinis2015-Molecule,kandala2017hardware}. Equally exciting are analog simulation experiments in these circuits, studying low disorder lattices~\cite{underwood2012low}, low-loss synthetic gauge fields~\cite{Roushan2016,owens2017quarter}, dissipative lattices~\cite{raftery2014observation,fitzpatrick2017observation} and many-body localization in disorder potentials~\cite{roushan2017spectral}. In the circuit platform, the particles that populate the system are microwave photons that are inevitably subject to intrinsic particle losses. Without an imposed chemical potential, the photonic system eventually decays to the vacuum state, naturally posing the challenge of how to achieve strongly-correlated matter in the absence of particle number conservation. To this end, dissipative preparation and manipulation of quantum states via tailored reservoirs have become an active area of research both theoretically and experimentally, where dissipative coupling to the environment serves as a resource~\cite{poyatos1996quantum,Plenio1999,Biella2017}. Such engineered dissipation has been employed to stabilize entangled states of ions~\cite{barreiro2011open}, single qubit states \cite{lu2017universal}, entangled two qubit states~\cite{shankar2013autonomously}, and holds promise for autonomous quantum error correction~\cite{kapit2015passive,kapit2016hardware,albert2018}.

Here, we present a circuit platform for exploration of quantum matter composed of strongly interacting microwave photons and employ it to demonstrate direct dissipative stabilization of a strongly correlated phase of photons. Our scheme~\cite{ma2017autonomous} builds upon and simplifies prior proposals~\cite{kapit2014,hafezi2015,lebreuilly2016towards,Lebreuilly2017}, and is agnostic to the target phase so long as it is incompressible and exhibits mobile quasi-holes.

To understand the protocol, illustrated in Fig.~\ref{fig:0}, consider a target ground state, comprising $N_0$ photons, that is spectrally gapped from excited states with the same particle number with a many-body gap $\Delta_{\textrm{mb}}$. It must additionally be incompressible with respect to change in particle number, in the sense that inserting each of the first $N_0$ particles requires about the same energy, while adding an $(N_0+1)$th particle requires an energy different by the compressibility gap $\Delta_{\textrm{comp}}$. Using a combination of coherent drive and engineered dissipation, we irreversibly inject particles into the system near the energy (per particle) of the target state. As long as the target state has good wavefunction overlap with both the initial state (e.g. the vacuum $N=0$) and the locally injected particles, the system will be continuously filled up to the target state, at which point further addition of particles is energetically suppressed by $\Delta_{\textrm{comp}}$. Generically, the injected particles will \textit{order} in the strongly correlated phase under the influence of the underlying coherent interactions, geometries or topological properties present in the many-body system. Population of other excited states is highly suppressed by spectral gaps, and further made short-lived by engineering an energy dependent loss that couples only excited manifolds to the environment. The balance of particle-injection and loss that is built into the system provides the autonomous feedback that populates the target many-body state, stabilizing it against intrinsic photon loss or accidental excitation.

We realize irreversible particle insertion by coherent injection of pairs of particles into a ``collider'', in which they undergo elastic collisions wherein one particle dissipates into an engineered cold reservoir while the other enters the many-body system; loss of the former particle makes this otherwise coherent process irreversible, permanently inserting the latter into the system. Prior experiments demonstrating ``optical pumping'' into spectrally resolved few-body states~\cite{hacohen2015cooling} relied upon excited-state symmetry to achieve state-dependent dissipation; here we employ energy-dependent photon loss to shed entropy, a new approach with broad applicability.

In Sec.~\ref{sec:BH_circuit}, we introduce and characterize the photonic Bose-Hubbard circuit; in Sec.~\ref{sec:stabilize_onesite} we describe and explore an isolated dissipative stabilizer for a single lattice site; finally, in Sec.~\ref{sec:stabilize_mott} we couple the stabilizer to the Bose-Hubbard circuit, realize the stabilization of a Mott insulating phase, and investigate the fate of defects in the stabilized Mott phase.

\begin{figure}[htbp]
    \includegraphics[width=1.0\columnwidth]{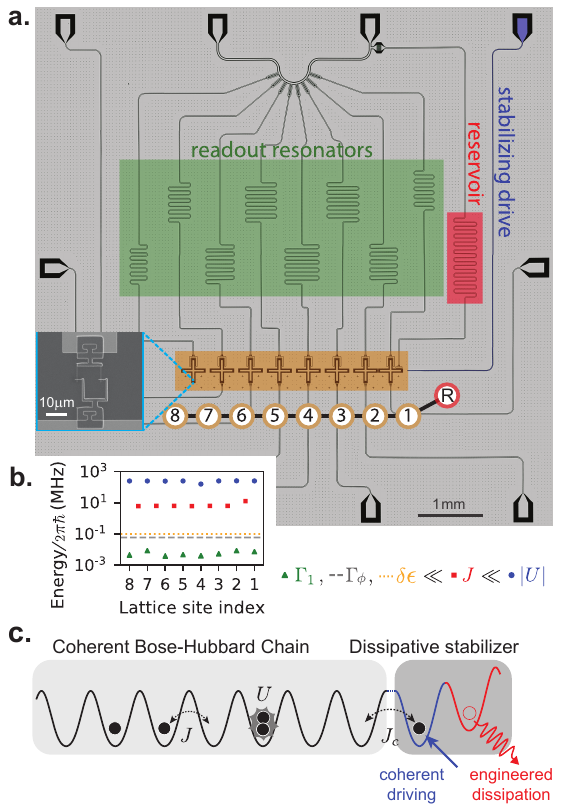}
    \caption{\textbf{Building a Bose-Hubbard lattice in superconducting circuits.}
    \textbf{(a) The circuit lattice} Optical image of the sample, showing an array of superconducting transmon qubits (yellow, $Q_1$-$Q_8$) that constitute the lattice sites of the Bose-Hubbard chain. Capacitive coupling between qubits leads to nearest neighbor tunneling $J$, while the transmon anharmonicity gives the on-site interaction $U$.  Lattice-site frequencies are dynamically tunable by independent flux bias lines. Individual readout resonators (green) enable site-resolved occupancy readout via a common transmission line. Additionally, site $Q_1$ is tunnel coupled to a lossy resonator (red) that acts as the cold reservoir for the dissipative stabilization. Charge excitation of the lattice sites is realized by driving via the readout transmission line, and a separate stabilization drive line (blue) couples only to site $Q_1$.
    \textbf{Inset}: Close-up SEM image of the transmon qubit, showing the bottom of the cross shaped capacitor pad and the SQUID loop. For details of the sample parameters and fabrication, see SI~\ref{a1}.
    \textbf{(b) Energy scales} The measured on-site interactions $U$, tunneling rates $J$, single photon losses $\Gamma_1$, dephasing rate $\Gamma_\phi$, and on-site disorder $\delta\epsilon$ are plotted, demonstrating a highly coherent, low-disorder Bose-Hubbard lattice in the strongly interacting regime. 
    \textbf{(c) Lattice illustration} The circuit sample corresponds to a coherent Bose-Hubbard chain (light shade), tunnel coupled at one end to the dissipative stabilizer (darker shade) with coupling $J_c$. Shown here is a particular implementation of the stabilizer using one transmon (blue) and the reservoir (red).
    }
    \label{fig:1_Schematic}
\end{figure}

\section{Building a Bose-Hubbard Circuit}
\label{sec:BH_circuit}

Figure~\ref{fig:1_Schematic}\textbf{a} shows our circuit, realizing a one-dimensional Bose-Hubbard lattice for microwave photons, with a Hamiltonian given by:

\begin{equation*}
\mathcal{H}_{\text{BH}}/\hbar = -\sum_{<ij>}{J_{ij} a_i^\dagger a_j}+\frac{U}{2}\sum_i{n_i(n_i-1)} + \sum_i \epsilon_i n_i
\end{equation*}

Here $a_i^\dagger$ is the bosonic creation operator for a photon on site $i$, $J_{ij}$ is the nearest neighbor tunneling rate, $U$ is the on-site interaction, $\epsilon_i$ is the local site energy, and $\hbar$ is the reduced Planck constant. An array of eight transmon qubits~\cite{koch2007charge} constitute the lattice sites of the one-dimensional lattice. Each transmon acts as a non-linear resonator, where a Josephson junction acts as a non-linear inductor with Josephson energy $E_J$, in parallel with a cross-shaped metal capacitor with charging energy $E_C = e^2/2C_\Sigma $, where $e$ is the electron charge and $C_\Sigma$ the total capacitance of the transmon. The lattice site has a frequency for adding only one photon of $\epsilon = \omega_{01} \approx \sqrt{8E_J E_C}$. Adding a second photon requires a different amount of energy with the difference given by the anharmonicity of the transmon $U = \omega_{12}-\omega_{01} \approx -E_C$. Thus $U$ is the effective two-body on-site interaction for photons on a lattice site. By using tunable transmons where two junctions form a SQUID loop, we control the effective $E_J$ and thus the site energy by varying the magnetic flux through the loop, achieved via currents applied to individual galvanically coupled flux-bias lines. Neighboring lattice sites are capacitively coupled to one another, producing fixed nearest neighbor tunneling $J_{ij}$. 

Each lattice site (transmon) is capacitively coupled to an off-resonant coplanar waveguide readout resonator, enabling site-by-site readout of photon number occupation via the dispersive shift of the resonator. The readout resonators are capacitively coupled to a common transmission line to allow simultaneous readout of multiple lattice sites and thereby site-resolved microscopy of the lattice. Main contributions to the readout uncertainty are Landau-Zener transfers between neighboring sites during the ramp to the readout energy, and errors from the dispersive readout (SI.~\ref{SI:readout+errorbar}). The readout transmission line also enables charge excitation of all lattice sites.

Site $Q_1$, at one end of the lattice, is coupled to another resonator which serves as a narrow band reservoir used for the dissipative stabilization. The reservoir is tunnel-coupled to $Q_1$ with $J_{R1} = 2\pi\times 16.3$\,MHz and has a linewidth $\kappa_R = 2\pi\times 9.5$\,MHz obtained by coupling to the $50$\,$\Omega$ environment of the readout transmission line. An additional drive line is capacitively coupled to $Q_1$ at the end of the lattice to allow direct charge excitation of only $Q_1$, used for the dissipative stabilization.

We employ transmon qubits with a negative anharmonicity $U \sim 2\pi\times -255$\,MHz corresponding to strong attractive interactions, and an on-site frequency tuning range of $\omega_{01} \sim 2\pi\times (3.5-6.0)$\,GHz with a tuning bandwidth of $250$\,MHz. We measure nearly-uniform tunneling rates of $\approx 2\pi\times 6.25$\,MHz for $J_{23}$ to $J_{78}$, and $J_{12} = 2\pi\times 12.5$\,MHz designed to optimize the dissipative stabilization. Beyond-nearest-neighbor-tunneling due to residual capacitance between qubits is suppressed by an order of magnitude. The excited-state structure of the transmon gives rise to effective on-site multi-body interaction terms that are irrelevant for experiments in this work, where the on-site occupancies are predominantly confined to $n=0,1,2$.

We measure single photon relaxation times $T_1 \sim 30$\,$\mu$s and dephasing times $T_2^* \sim 3$\,$\mu$s for the lattice sites (see SI~\ref{SI:sub-coherences}), corresponding to a single photon loss rate $\Gamma_1 = 1/T_1 \sim 2\pi\times 5$\,kHz and on-site frequency fluctuation $\Gamma_\phi = 1/T_2^* \sim 2\pi\times 50$\,kHz. We have thus realized a highly coherent photonic Bose-Hubbard lattice in the strongly interacting regime $|U| \gg J \gg \Gamma_1,\Gamma_\phi$, as shown in Fig.~\ref{fig:1_Schematic}\textbf{b}.
The on-site frequency disorder is another crucial characteristics that should be compared with the other energy scales of the lattice: tunneling, interaction, and more generally the many-body gap of the state being studied. We achieve on-site disorder $\delta\epsilon = \delta\omega_{01}\lesssim 2\pi\times 100$\,kHz, well below both $J$ and $U$, where $U$ is also the approximate excitation gap of the Mott state in the strongly interacting regime~\cite{greiner2002quantum}. Currently, dephasing $\Gamma_\phi$ is limited by electronic noise on the flux bias while disorder $\delta\epsilon$ is limited by precision of the flux bias calibration (SI.~\ref{SI:flux_control}); neither impacts present experiments.

\section{Dissipative stabilization of a single lattice site}
\label{sec:stabilize_onesite}

\begin{figure*}[t]
\includegraphics[width=1.0\textwidth]{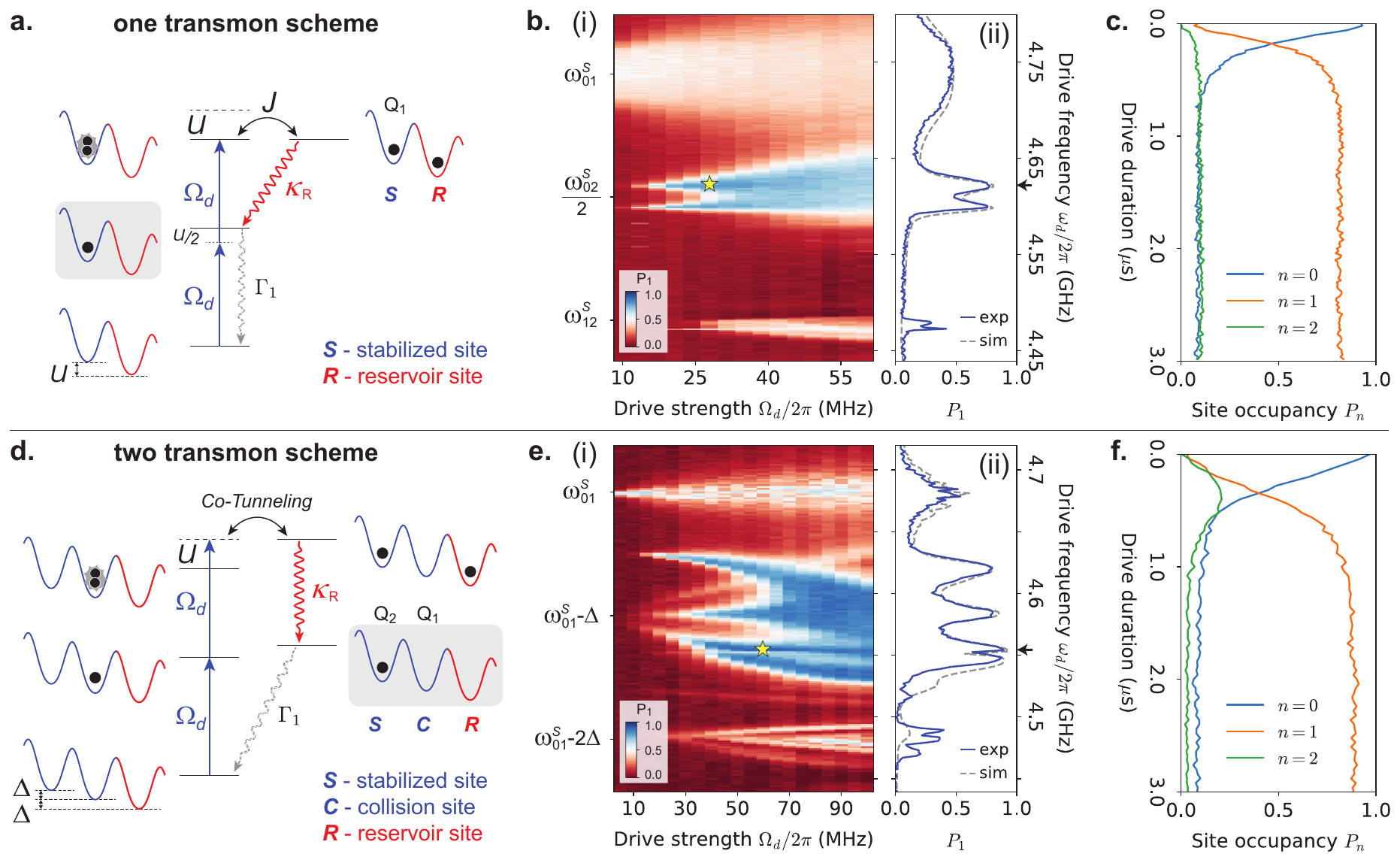}
\caption{\textbf{Dissipative stabilization of a single lattice site.} 
To stabilize the population of a single lattice site, we explore two different approaches: \textbf{(a) One transmon scheme:} Making use of the on-site interaction, we charge-drive the 2-photon transition from $n=0$ to $n=2$ at frequency $\omega_{d} = \omega_{02}/2$, off resonant from the $n=1$ state by $U/2$. The $n=1$ to $n=2$ transition is resonantly coupled to a lossy site (reservoir) at frequency $\omega_R = \omega_{21}$ to enhance $\Gamma_{2\rightarrow 1}$. This causes the stabilized site to always return back to the $n=1$ state, shaded in gray. Solid (dashed) black lines indicate energy levels of the illustrated states in the presence (absence) of interaction $U$.
\textbf{(d) Two transmon scheme:} Two transmon sites and the lossy site are detuned in a Wannier-Stark ladder configuration (with detuning $\Delta$) to stabilize a single lattice site. Charge-driving resonantly at the energy of the collision site coherently injects pairs of photons, which collide elastically and split, with one going into the stablized site and the other to the reservoir. When the photon in the reservoir dissipates, the stabilized site remains in the $n=1$ state.
\textbf{(b)(e) Single site stabilizer fidelity.}  (i) The stabilizer fidelity $P_1$ is measured as a function of the driving frequency $\omega_d$ and driving strength $\Omega_d$ after a driving duration of $3$\,$\mu$s.
The highest observed fidelities are $0.81(\pm 0.01)$ for the one transmon scheme, and $0.89(+0.04/\text{\textminus}0.01)$ for the two transmon scheme, limited predominantly by finite thermal population in the reservoir. In (ii) we plot the fidelities at the optimal $\Omega_d$ and compare to numerics, showing quantitative agreement. \textbf{(c)(f) Stabilizer filling dynamics.} We measure the on-site occupancy of the stabilized site as a function of the duration of stabilization drive, at the optimal driving parameters (indicated as stars and arrows in \textbf{(b)(e)}). The non-zero $P_1$ at $t=0$ is a result of the equilibrium qubit temperature. All data points in the paper are averages of $\sim (5000-8000)$ independent runs of the experiment performed at a repetition rate of 4\,kHz. Typical errorbars in the stabilization fidelity near optimal parameter-regime are $(+2/\text{\textminus} 1)\%$ , dominated by systematic uncertainties from unwanted population transfer during readout (SI.~\ref{SI:readout+errorbar}).
}
\label{fig:3_onesite}
\end{figure*}

Before examining the more complicated challenge of stabilizing a Bose-Hubbard chain, we consider the following simpler question: how do we stabilize a single lattice site with exactly one photon in the presence of intrinsic single photon loss? A continuous coherent drive at $\omega_{01}$ can at best stabilize the site with an average single-excitation probability $P_1=0.5$ in the steady state, where the $n=2$ state remains un-populated due to strong interactions making the drive off-resonant for $1\rightarrow 2$ transition. To stabilize in the $n=1$ state, one could implement a discrete feedback scheme where the state of the site is continuously monitored and whenever the occupation decays from $n=1$ to $n=0$, a resonant $\pi$ pulse injects a single photon into the site. Such active feedback requires constant high efficiency detection, fast classical control, and works only for simple separable states. Here we explore ways to implement the stabilization autonomously using an engineered reservoir. The autonomous approach has the required feedback built into the driven-dissipative Hamiltonian, allowing the preparation of many-body states with strong and even unknown correlations.

This idea of autonomous stabilization is akin to inverting atoms in a laser or optical pumping schemes prevalent in atomic physics: a coherent optical field continuously drives an atom from the ground state to a short-lived excited state that rapidly decays to a long-lived target state. In the transmon, this means making one photon significantly shorter lived than the other; to this end it is helpful to be able to distinguish them, e.g. by different spatial wavefunctions or different energies. We take the latter route, harnessing on-site interactions and elastic site-changing collisions to allow the coherent field to add pairs of photons with different energies, and the narrow band reservoir to provide an energy dependent loss into which the lattice-site's entropy is shed, stabilizing the site into the $n=1$ state.

\begin{figure*}
\includegraphics[width=1.0\textwidth]{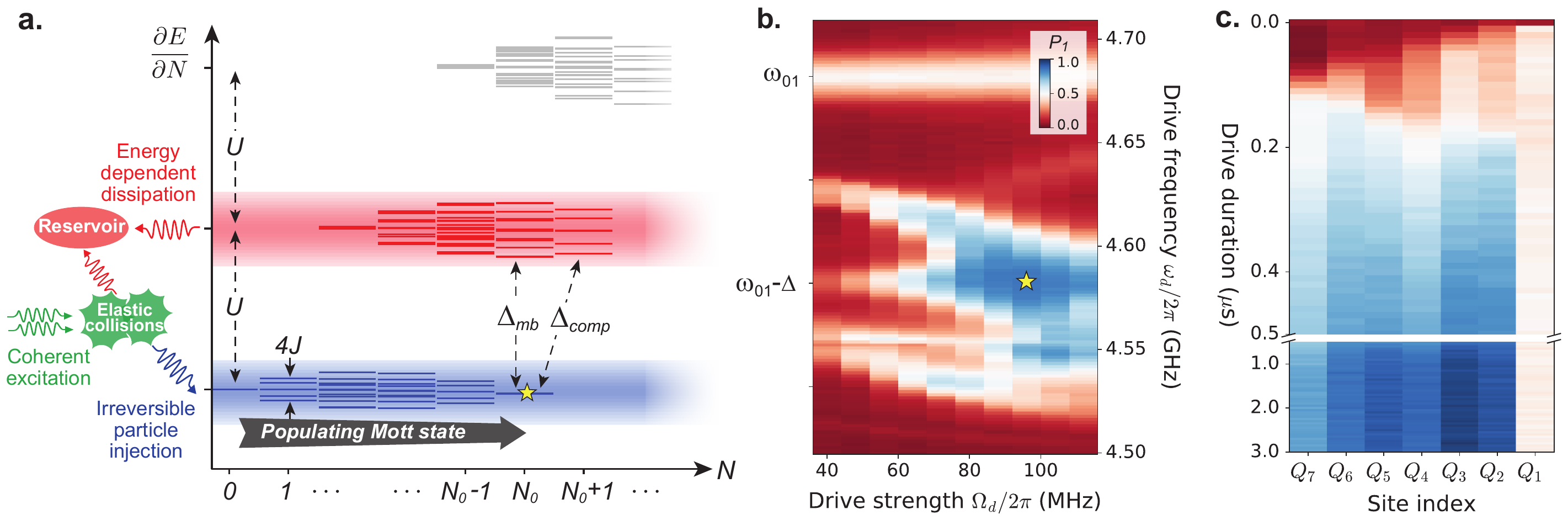}
\caption{\textbf{Dissipative stabilization of a Mott insulator.} \textbf{(a) Schematics:} Many-body spectrum of a strongly interacting Bose-Hubbard chain showing bands separated by the interaction $U$. The continuous coherent driving provides particle pairs that undergo elastic collisions where one dissipates into the reservoir, and the other irreversibly injected into the lowest band of the lattice (blue shade). The lattice is filled up to and stabilized in the $n=1$ Mott state (gray arrow) which is both gapped ($\Delta_{\textrm{mb}}$) and incompressible ($\Delta_{\textrm{comp}}$). The reservoir also acts as an dissipation channel for any unwanted excitations to higher energy states (red shade). For large $|U|/J$, the setup applies to both positive or negative $U$.
\textbf{(b) Mott fidelity:} We take the stabilizer in the ``two transmon scheme'' and resonantly couple it to the end of a homogeneous 5 site Bose-Hubbard chain. The measured steady state $P_1$ averaged over sites $Q_2$ to $Q_7$ is shown as a function of the stabilization drive frequency and drive strength, displaying same qualitative features as when stabilizing a single site (Fig.~\ref{fig:3_onesite}\textbf{e}). We achieve a maximum average Mott fidelity of $0.88(+0.03/\text{\textminus}0.01)$ when driving on resonance at the collision site with $\Omega_d = 2\pi\times 95$\,MHz (indicated by star). 
\textbf{(c) Mott filling dynamics:} Starting with an empty lattice, we plot $P_1$ on each site as a function of the stabilization drive duration at the optimal driving parameters. \textbf{(b)} and \textbf{(c)} share the same colorbar scale. At short times, particles are injected continuously into the lattice from the stabilizer and display a light-cone-like ballistic transport of population across the lattice, and a single reflection off of the far end of the lattice. The steady state Mott filling is reached in $\sim 0.8$\,$\mu$s.
}
\label{fig:4_mott}
\end{figure*}

We implement two different schemes for stabilizing a single lattice site. 
In the ``one transmon'' scheme (Fig.~\ref{fig:3_onesite}\textbf{a}), akin to~\cite{leek2009using}, we make use of the on-site $n=2$ state and drive a 2-photon transition from $n=0$ to $n=2$ at frequency $\omega_{d} = (\omega_{01} + \omega_{12})/2$ and single photon Rabi rate $\Omega_d$, off resonant from the $n=1$ state by $U/2$. The $2\rightarrow 1$ photon loss is realized by coupling the stabilized site to the lossy site ($R$) at frequency $\omega_R = \omega_{21}$. The optimal stabilization fidelity $P_1$ (probability of having on-site photon occupancy $n=1$) arises from a competition between the coherent pumping rate and various loss processes: at low pumping rates, the photons are not injected fast enough to compete with the 1-photon loss $\Gamma_1$; at high pumping rates, the lossy site cannot shed the excess photons fast enough and the fidelity is limited by off-resonant coherent admixtures of zero- and two- photon states. The theoretically predicted single site infidelity $(1-P_1)$ for optimal lossy channel and driving parameters scales as $(\Gamma_1/|U|)^{1/2}$~\cite{ma2017autonomous}. The sign of interaction $U$ does not affect the physics of the experiments described in this paper, as the engineered reservoir is narrow-band and the lattice remains in the strongly interacting regime.

The measured steady-state stabilization fidelity using the ``one transmon'' scheme is shown in Fig.~\ref{fig:3_onesite}\textbf{b} as a function of the driving frequency and strength. Driving the stabilized site resonantly at $\omega_d \approx \omega_{01} = 2\pi\times 4.738$\,GHz gives an on-site population that saturates at $P_1 \leq 0.5$ as expected. Near $\omega_{d}= (\omega_{01} + \omega_{12})/2 = 2\pi\times 4.610$\,GHz we observe the single site stabilization and the fidelity increases with driving strength until reaching an optimal value $P_1 = 0.81(\pm 0.01)$ at $\Omega_d = 2\pi\times 28$\,MHz after which the fidelity drops. The split peaks result from resonant coupling between the lossy resonator and the stabilized site, $\omega_R=\omega_{12}$, giving a frequency splitting of $\sqrt{2} J_{R1} \approx 2\pi\times 23$\,MHz when driving the 2-photon transition. The measured data at the optimal $\Omega_d$ is plotted in the vertical panel, showing quantitative agreement with a parameter-free numerical model (SI.~\ref{SI:numerics-single_site_stab}). The observed stabilization fidelity is primarily limited by thermal population in the cold-reservoir $n_{\mathrm{th}}^R = 0.075$ which re-enter the stabilized site (SI.~\ref{SI:sub-n_thermal}). In Fig.~\ref{fig:3_onesite}\textbf{c} we show the filling dynamics of the stabilization process, plotting on-site occupancy of the stabilized site versus the duration of the stabilization drive with the optimal driving parameters (star, arrow in Fig.~\ref{fig:3_onesite}\textbf{b}). The single site is filled in about $0.8$\,$\mu$s (with a fitted exponential time constant of $0.19$\,$\mu$s), in agreement with numerical simulations. The finite $P_1$ at time $t=0$ arises from finite qubit temperature in the absence of driving.

In the ``two transmon'' scheme, we employ two transmon lattice sites (the ``stabilized site (\textit{S})'' and the ``collision site (\textit{C})'') and the lossy resonator (the ``reservoir (\textit{R})'') in a Wannier-Stark ladder configuration (Fig.~\ref{fig:3_onesite}\textbf{d}). The middle collision site is placed energetically between the stabilized and the reservoir sites, detuned from each by $\Delta$, allowing us to drive at the collision site frequency $\omega_{01}^C$ and induce elastic collisions that put one photon each into the stabilized site and lossy resonator ($2\times\omega_{01}^C \rightarrow \omega_{01}^S + \omega_R$). The photon in the reservoir site is quickly lost, leaving the stabilized site in the $n=1$ state. This scheme resembles evaporative cooling employed in ultracold atom experiments where an RF knife in a magnetic trap provides an energy-dependent loss at the edge of the quantum gas, elastic collisions cause one particle to gain energy and spill out of the trap, while the other is cooled~\cite{anderson1995observation}. Compared to the ``one transmon'' scheme, the ``two transmon'' scheme adds an additional degree of freedom, making it possible to separate the effective pumping rate from the detuning, allowing for better stabilization performance where the optimal infidelity scales as $(\Gamma_1/|U|)^{2/3}$~\cite{ma2017autonomous}. In addition, the stabilized site is not driven directly in the ``two transmon'' scheme, thus avoiding infidelities from off-resonant population of higher transmon levels.

The measured steady state fidelity of the stabilized site in the ``two transmon'' scheme is shown in Fig.~\ref{fig:3_onesite}\textbf{e} with $\Delta= 2\pi\times 100$\,MHz chosen for optimal fidelity. The expected stabilization peak at the collision site frequency $\omega_d = \omega_{01}^C = \omega_{01}^S-\Delta$ is observed, accompanied by other features with high $P_1$ from higher-order collision processes~\cite{ma2011photon} (See SI.~\ref{SI:numerics-single_site_stab} for details). The measured optimal single site stabilizer fidelity is $P_1 = 0.89(+0.04/\text{\textminus}0.01)$ at $\Omega_d = 2\pi\times 60$\,MHz, $\omega_d = 2\pi\times 4.555$\,GHz. Both the measured steady-state fidelity and the stabilizer dynamics (Fig.~\ref{fig:3_onesite}\textbf{f}) are in quantitative agreement with numerical simulation, with the highest observed fidelity primarily limited by reservoir thermal population. Notice that compare to the ``one transmon'' scheme, the ``two transmon'' scheme yields higher fidelities and does so over a broader parameter range (for example the $\omega_d \sim \omega_{01}^S - \Delta$ peak at higher $\Omega_d$); it is thus better suited to stabilize many-body states as described in the next section, where rapid refilling over a finite density of states is required.

\section{Stabilization of a Mott insulator}
\label{sec:stabilize_mott}

Having demonstrated the ability to stabilize a lattice site with a single photon, we now employ it to stabilize many-body states in the Bose-Hubbard chain. The single stabilized site acts like a spectrally narrow-band photon source that is continuously replenished. Photons from it sequentially tunnel into and gradually fill up the many-body system until adding further photons requires an energy different from that of the source, resulting in ordering of the photons by their strong coherent interaction into a Mott insulator with near-perfect site-to-site particle number correlation in the Hubbard chain~\cite{greiner2002quantum}.

In order for this stabilization method to work, the target phase must satisfy certain conditions, illustrated for the current system in Fig.~\ref{fig:4_mott}\textbf{a}. The phase should be incompressible with respect to particle addition~\cite{ma2017autonomous}: Once the target state is reached, the stabilizer should be unable to inject additional photons into the system; at the same time, when a photon is lost from the target state (due to decay into the environment), the stabilizer must refill the hole defect efficiently. This requires that the hole- and particle- excitation spectra of the target state be spectrally separated (with gap $\sim \Delta_{\textrm{comp}}$). In addition, when refilling a hole, we must avoid driving the system into excited states with the same number of photons as the target state-- requiring the target phase to exhibit a many-body gap $\Delta_{\textrm{mb}}$. The stabilizer, as a continuous photon source, thus needs to be narrow-band compared to both the many-body gap and the gap between the hole states and the particle states, but sufficiently broad-band to spectroscopically address all hole states. The performance of the many-body stabilizer is then determined by how efficiently the hole defects in the many-body state can be refilled-- a combined effect of the repumping rate of the single site stabilizer at energy $\epsilon_k$ (where $k$ is the quasi-momentum of the hole) and the wavefunction overlap between the defect state and the stabilizer site.

\begin{figure}
\includegraphics[width=1.0\columnwidth]{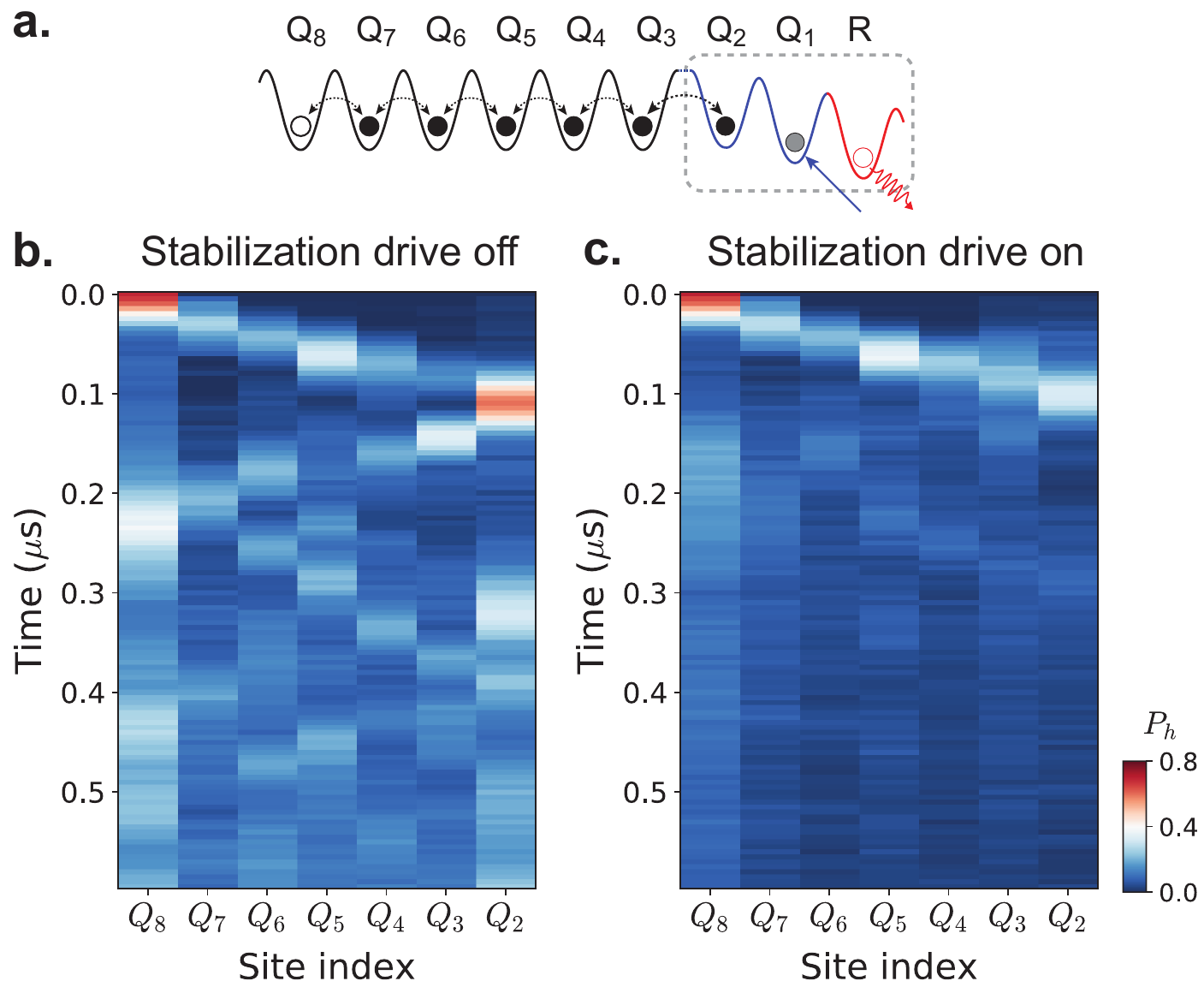}
\caption{\textbf{Dynamics of a hole defect in the Mott insulator.} \textbf{(a) Schematics} To explore thermalization dynamics near steady state, we prepare a Mott insulator with a localized hole defect at the end of the lattice opposite the stabilizer and observe the time dynamics of the chain. \textbf{(b)} We plot the excess hole density $P_h$ as a function of evolution time in the absence of the stabilization drive. The hole propagates on top of the otherwise filled Mott insulator as the coherent quantum walk of a free quasi-particle. \textbf{(c)} If the stabilization drive is on during the evolution, the hole shows the same ballistic propagation in the lattice until it reaches the stabilizer, where the defect is efficiently refilled.
}
\label{fig:5_holedyn}
\end{figure}

We tunnel-couple the demonstrated single site stabilizer to one end of the Bose-Hubbard chain, and attempt to stabilize the $n=1$ Mott insulator of photons. The Mott state is a gapped ground state~\cite{greiner2002quantum} that satisfies the incompressibility requirements~\cite{gemelke2009situ}. The many-body gap is set by the cost to create doublon-hole excitations on top of the Mott state $\approx U$. Particle-like excitations are gapped by the strong interaction ($\approx U$ for $n=1$ Mott state), while the hole excitations follow the single particle dispersion with energies lying in a band of $|\epsilon_k|\leq 2J$ in the one-dimensional lattice, providing clear spectral separation in the Mott limit ($|U|\gg J$). For a homogeneous lattice, all hole eigenstates are delocalized across the lattice, making it possible to employ a single stabilizer at one end of the chain. The amplitude of the defect state wavefunctions at the stabilizer can be adjusted via the coupling between the chain and the stabilizer $J_c$. Here we attach an additional 5 site chain ($Q_3 - Q_7$) to a ``two transmon'' stabilizer which stabilizes $Q_2$. All lattice sites are tuned to the same energy as $Q_2$. The coupling between the stabilizer and the rest of the chain is $J_c \approx J_{\textrm{chain}} \sim 2\pi\times 6.25$\,MHz.

In Fig.~\ref{fig:4_mott}\textbf{b}, we plot the measured steady state Mott fidelity ($\langle P_1 \rangle$, chain-averaged over sites $Q_2 - Q_7$) as a function of the stabilization drive frequency and strength, after a driving duration of $5 \mu$s. The optimal Mott fidelity of $0.88(+0.03/\text{\textminus}0.01)$ is achieved by driving at $\omega_d = \omega_{01}-\Delta$ with $\Omega_{d} = 2\pi\times 96$\,MHz, demonstrating a dissipatively stabilized photonic Mott insulator in which the on-site number fluctuations are strongly suppressed. The observed defects within the chain are predominantly holes ($P_0\approx 0.10$), with very low doublon probabilities ($P_2\approx 0.02$) (SI.\ref{SI:readout+errorbar}). Ignoring small site-to-site variations in the Mott fidelity, we obtain the on-site number fluctuation of the Mott state $\delta n \equiv \sqrt{\langle n^2\rangle-\langle n\rangle^2} = 0.34(+0.01/\text{\textminus}0.05)$; or a configuration entropy of $s \equiv - k_B \sum_n P_n \ln{P_n} = 0.42(+0.04/\text{\textminus}0.12)\, k_B$ per site, where $k_B$ is the Boltzmann constant. Figure~\ref{fig:4_mott}\textbf{b} shows qualitatively the same features as the single site stabilization in Fig.~\ref{fig:3_onesite}\textbf{d}. Near $\omega_d\approx\omega_{01}^C$, the single particle stabilizer performance is robust over variations in both (1) drive detuning, which gives good energetic overlap with the hole defect states of the Mott phase that span a frequency range of $4J \approx 2\pi\times 25$\,MHz; and (2) drive strength, which provides the high repumping rates necessary to fill the whole lattice without sacrificing stabilizer fidelity.
For larger lattices with a single stabilizer site, the stabilization performance will eventually be limited by the reduced refilling rate for the increasing number of sites/modes, and by disorder induced localization that inhibits the effective refilling of defects away from the stabilizer site. Multiple stabilizers may be used to circumvent such limitations, as envisioned in proposals~\cite{Verstraete2009, kapit2014} where \textit{each} lattice site is coupled to a driven-dissipative bath.

In Fig.~\ref{fig:4_mott}\textbf{c} we plot the time dynamics of all lattice sites in the Hubbard chain as the Mott state is filled from vacuum, at the optimal driving parameter (indicated with yellow star in Fig.~\ref{fig:4_mott}\textbf{b}.). The initial filling dynamics reveal near-ballistic propagation of injected photons after they enter the lattice from the stabilizer, consistent with the dispersion of a localized wavepacket continuously injected at the stabilized site that undergoes quantum tunneling in the lattice. We observe light-cone-like transport~\cite{Cheneau2012} at a speed of approximately $2J\approx 78\,\textrm{sites}/{\mu\textrm{s}}$ ($13\,\textrm{ns}/\textrm{site}$). In comparison, the single site refilling time at these Mott driving parameters (Fig.~\ref{fig:3_onesite}\textbf{e}) is about $45\,\textrm{ns}$ and remains relatively uniform over the lattice bandwidth of $4J$.
It is natural to ask how our dissipatively prepared Mott state relates to the corresponding one in an isolated system at equilibrium. Fundamentally, this is a question of the timescales between thermalization within the system and interaction with the reservoir. In future work, we can measure density-density correlations or entanglement~\cite{Islam2015} to compare the dissipatively prepared Mott insulator to an equilibrium Mott insulator at finite temperature, as well as investigate how the stabilized wavefunctions vary with distance from the stabilizer. 

Finally, we examine the near-steady-state dynamics of the stabilized chain by preparing a single defect and watching it refill (Fig.~\ref{fig:5_holedyn}). We begin by preparing the dissipatively stabilized Mott insulator in $Q_2 - Q_7$ with $Q_8$ sufficiently energetically detuned that it remains empty. $Q_8$ is then rapidly tuned to resonance with the rest of the lattice, and the population of holes (excess $n=0$ population compared to the steady state Mott, $P_h = P_0 - P_0^{\textrm{Mott}}$) is measured across the chain after a variable evolution time. In the absence of the stabilization-drive during the evolution of the hole (Fig.~\ref{fig:5_holedyn}\textbf{b}), we observe the coherent propagation of the hole defect (consistent with theory, see SI~\ref{SI:numerics-hole_dynamics}). The wavefront traverses the lattice at a speed of $2J$ at short times, while at longer times we observe complex structures emerge due to coherent interference of multiple reflections off the edges of the lattice. On the other hand, when the stabilization drive remains on during the evolution of the hole defect (Fig.~\ref{fig:5_holedyn}\textbf{c}), we observe similar initial ballistic propagation until the defect reaches the stabilizer, where the hole defect is immediately filled. Note that the many-body filling front in Fig.~\ref{fig:4_mott}\textbf{c} is essentially as fast as the single hole propagation shown in Fig.~\ref{fig:5_holedyn}\textbf{c}.

\section{Conclusions}
\label{sec:conclusions}

We have constructed a Bose-Hubbard lattice for microwave photons in superconducting circuits. Transmon qubits serve as individual lattice sites where the anharmonicity of the qubits provides the strong on-site interaction, and capacitive coupling between qubits leads to fixed nearest-neighbor tunneling. The long coherence times of the qubits, together with the precise dynamical control of their transition frequencies, make this device an ideal platform for exploring quantum materials. Using readout resonators dispersively coupled to each lattice site, we achieve time- and site- resolved detection of the lattice occupancy. Frequency multiplexed simultaneous readout of multiple lattice sites~\cite{Jeffrey2014Readout} could be implemented in future experiments to enable direct measurement of entanglement and emergence of many-body correlations.

We further demonstrate a dissipative scheme to populate and stabilize gapped, incompressible phases of strongly interacting photons-- employed here to realize the first Mott insulator of photons. The combination of coherent driving and engineered dissipation creates a tailored environment which continuously replenishes the many-body system with photons that order into a strongly correlated phase and acts as an entropy dump for any excitation on top of the target phase. The dissipatively prepared incompressible phases can serve as a starting point for exploring other strongly-correlated phases via coherent adiabatic passages, including compressible ones. The latter are also proposed to be accessible directly with dissipative preparation~\cite{hafezi2015, Lebreuilly2017}.

This platform opens numerous fascinating avenues for future exploration: What is the optimal spectral and/or spatial distribution of engineered reservoirs? How does this depend upon the excitation spectrum of the isolated model under consideration? How do the equilibrium properties of the dissipatively stabilized system relate to those of the isolated system? How do higher-order correlations emerge and thermalize? What are the thermodynamic figures of merit for the reservoir and its coupling to the system?

Finally, our results provide an exciting path towards topologically ordered matter using related tools, e.g. the creation of fractional quantum Hall states of photons~\cite{umucalilar2012fractional,anderson2016engineering} in recently realized low-loss microwave Chern insulator lattices~\cite{owens2017quarter}. The unique ability in circuit models to realize exotic real-space connectivity~\cite{ningyuan2015time} further suggests the possibility of exploration of topological fluids on reconfigurable higher-genus surfaces-- a direct route to anyonic braiding~\cite{barkeshli2012topological}.\\

\section*{Acknowledgements}
\label{sec:acknowledgements}
We would like to thank Mohammad Hafezi and Andrew Houck for fruitful discussions. This work was supported by Army Research Office grant W911NF-15-1-0397; and by the University of Chicago Materials Research Science and Engineering Center (MRSEC), which is funded by National Science Foundation (NSF) under award number DMR-1420709. D.I.S. acknowledges support from the David and Lucile Packard Foundation; R.M. acknowledges support from the MRSEC-funded Kadanoff-Rice Postdoctoral Research Fellowship; C.O. is supported by the NSF Graduate Research Fellowships Program. This work also made use of the Pritzker Nanofabrication Facility at the University of Chicago, which receives support from NSF ECCS-1542205.

\medskip
\noindent \textbf{Data Availability}
All experimental data and numerical simulations presented in this manuscript are available upon request.

\medskip
\noindent \textbf{Author Contributions}
R.M., B.S., C.O., J.S. and D.I.S. designed and developed the experiments. R.M. and B.S. performed the device fabrication, measurements and analysis, with assistance from N.L. and Y.L. All authors contributed to the preparation of the manuscript.



\onecolumngrid
\pagebreak

\section*{A dissipatively stabilized Mott insulator of photons}

\section*{Supplementary Information}
\appendix

\renewcommand{\thefigure}{S\arabic{figure}}
\renewcommand{\thetable}{S\arabic{table}}

\setcounter{figure}{0}
\setcounter{table}{0}

\section{Device Fabrication and Parameters}
\label{a1}

The superconducting circuit device is fabricated in a two step process: (1) Optical lithography defines the capacitor pads for the transmon qubits, co-planer waveguide resonators, all control and input/output lines (flux biases, charge drive, readout transmission line) and the perforated ground plane; (2) E-beam lithography defines the \textit{dc} SQUID loops and forms the Josephson junctions for the transmons.

The base layer is fabricated from $150$\,nm of Niobium, e-beam evaporated (at $0.9$\,nm/s) onto $450\,\mu$m thick $c$-plane sapphire substrate that has been annealed at $1500\,\celsius$ for $1.5$\,hrs. Optical lithography is performed with a direct pattern writer (Heidelberg MLA 150), followed by fluorine etching ($\textrm{SF}_6/\textrm{CHF}_3/\textrm{Ar}$) in a PlasmaTherm ICP etcher. This defines all patterns on the sample except the Josephson junctions and traces that form the SQUID loops.

Next we perform e-beam lithography using MMA-PMMA bilayer resist, written on a $30$\,keV FEI Quanta system with NPGS pattern generator. The $\textrm{Al}/\textrm{AlO}_x/\textrm{Al}$ junctions are e-beam evaporated in an angled evaporator (Plassys MEB550). Before Al deposition, we use Ar ion milling on the exposed Nb to etch away the Nb oxide layer in order to ensure electrical contact between the Nb and Al layers. The first layer of Al ($60$\,nm, deposited at $0.1$\,nm/s) is evaporated at an angle of $30^{\circ}$ to normal, followed by static oxidation in $O_2$ for $12$ minutes at $20$\,mBar. The second layer of Al ($150$\,nm, $0.1$\,nm/s) is then evaporated at $30^{\circ}$ to normal but orthogonal to the first layer in the substrate plane to form the junctions at the cross of the two layers. In order to reduce sensitivity to flux noise while retaining sufficient frequency tunability, the transmons have two asymmetric square-shaped junctions with sizes of $110$\,nm and $180$\,nm. The SQUID loop has a dimension of $10\,\mu\text{m}\times 15\,\mu$m. 

The device is then wire-bonded and mounted to a multilayer copper PCB with microwave launchers. The device chip is enclosed by a pocketed OFHC copper fixture which has been designed to eliminate all spurious microwave modes near or below the frequencies of interest.

The transmons have cross-shaped capacitors and total capacitance $C_\Sigma$ with $E_C \sim 2\pi\times 250$\,MHz and flux tunable $E_J \sim 2\pi\times (6-18)$\,GHz, corresponding to a tunable frequency of $\omega_{01} \sim 2\pi\times (3.5-6.0)$\,GHz. The capacitance between the neighboring transmons $c$ sets the nearest neighbor tunneling $J \approx \omega_{01} c/2C_\Sigma$. The lossy resonator, which serves as the reservoir for the dissipative stabilization, is a $\lambda/2$ co-planer waveguide resonator at $\omega_R = 2\pi\times 4.483$\,GHz, tunnel coupled to the end of the lattice ($Q_1$) via capacitive coupling to the transmon capacitor pad. It has a linewidth of $\kappa_R = 2\pi\times 9.5$\,MHz due to coupling of the other end of the resonator to the $50\,\Omega$ terminated environment of the readout transmission line (via an interdigitated capacitor). 

The readout resonators for the individual qubits are $\lambda/2$ co-planer waveguide resonators staggered in frequency with a spread of $200$\,MHz around $6.35$\,GHz, coupled to the common readout transmission line via parallel capacitors. The flux bias lines are galvanically coupled to the SQUID loops with a mutual inductance of $\sim 0.2$\,pH. Characterization of the crosstalk between flux lines is detailed below in Sec.~\ref{SI:flux_control}.

The stabilization drive line is coupled to $Q_1$ with a capacitance of $\approx 0.04$\,fF, while residual capacitive coupling to the next site ($Q_2$) is suppressed by a factor of $\geq 10$. The physical proximity of the stabilization drive line to the reservoir resonator on the chip leads to some direct coupling between the two; this coupling term has negligible impact on stabilizer performance since coherent population of the reservoir is strongly suppressed by the large detuning between drive and reservoir. 
A detailed characterization of the Bose-Hubbard parameters and the lattice readout is provided in the sections below, with a summary of the measured parameters listed in Table~\ref{tab:SI_coherences}.

\section{Fridge and Microwave Setup}
\label{SI:fridge_wiring}

\begin{figure}
\includegraphics[width=0.7\columnwidth]{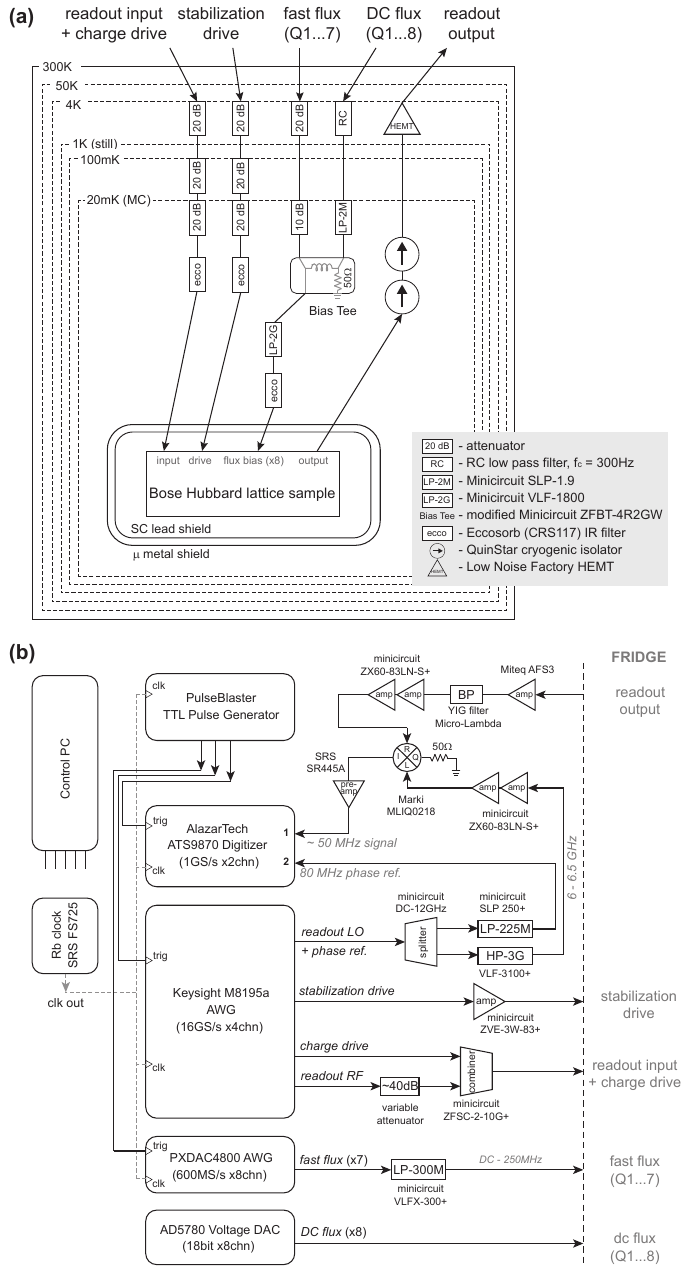}
\caption{Fridge wiring \textbf{(a)} and room temperature measurement setup \textbf{(b)}.}
\label{fig:SI_fridge-wiring}
\end{figure}

The packaged device is mounted using a machined high purity copper post to the base of a Bluefors dilution refrigerator at a nominal temperature of $20$\,mK. To provide additional shielding to radiation and external magnetic fields, the sample is enclosed in a thin high-purity copper shim shield, followed by a high-purity superconducting lead shield, and then two layers of cryo-compatible mu-metal shields (innermost to outermost). All shields are heat-sunk to the fridge base. 

Room temperature connections to the readout transmission line and the stabilization drive have $20$\,dB attenuation at each of the $4$\,K/$100$\,mK/$20$\,mK stages of the fridge, and Eccosorb (CRS-117) filters at base to block IR radiation. The readout line is also used to simultaneously charge drive all the qubits. The signal from output of the readout transmission line goes through two microwave isolators at base ($50$\,dB total isolation) prior to connection to a cryogenic HEMT amplifier (Low Noise Factory) at $4$\,K via a superconducting NbTi coax line.

There are a total of 7 on-chip flux lines attached to sites $Q_1$ to $Q_7$, allowing both \textit{dc} and fast tuning of the qubit frequencies. Site $Q_8$ furthest away from the thermalizer does not have an on-chip flux line. An external \textit{dc} coil ($20$ turn OFHC copper) is mounted $\sim 5$\,mm above the sample chip that allows simultaneous \textit{dc} tuning of all qubits, and together with the 7 on-chip lines, provides individual \textit{dc} frequency tuning of all 8 qubits. The \textit{dc} flux bias and fast flux bias are filtered separately before combined on bias-tees at base. The bias-tees allow pass-through of fast flux signal down to \textit{dc}. The \textit{dc} flux biases are generated by constant voltages sources (AD5780 DAC evaluation boards, 18 bit resolution, output range $\pm10$\,V). The fast flux tuning pulses are generated with PXDAC4800 Arbitrary Waveform Generators (AWGs) running at $600$\,MS/s and filtered to \textit{dc}$\,-\,250$\,MHz. 

All microwave signals for readout and charge drives are directly synthesized with a $64$\,GS/s Keysight M8195 AWG. The readout input signal (at $\omega_{\textrm{RF}}=\omega_{\textrm{read}}$) is attenuated and combined with the charge drive and sent into the readout transmission line. The stabilization drive is amplified and sent into the fridge separately. A fourth channel is used to generate simultaneously the local oscillator at $\omega_{\textrm{LO}} = \omega_{\textrm{read}}+50$\,MHz and a phase reference at $80$\,MHz synced to both the readout RF and LO.
The readout signal from the fridge goes through a low noise Miteq amplifier, filtered with a tunable $30$\,MHz bandwidth YIG filter, and further amplified. It is then mixed on a Marki IQ mixer with the readout LO and down-converted to $\sim 50$\,MHz. This heterodyne signal is further amplified and then recorded on an AlazarTech $1$\,Gs/s digitizer. We then perform ``digital homodyne'' to extract the two quadrature signals by extracting the amplitude of the sine and cosine components of the recorded trace. The $80$\,MHz signal is recorded simultaneously on the second channel of the digitizer and used as phase reference for the down converted heterodyne signal. This is to mitigate a small random timing jitter ($\sim 1-2$\,ns) between the digital trigger and the start of the Alazar card's acquisition.

All instruments for signal pulses and data acquisition are triggered with a PulseBlaster digital pulse generator, and clocked with a Rubidium atomic reference (SRS FS725).

The cryogenic microwave setup and the wiring of room temperature control/readout instrumentation are shown in Fig.~\ref{fig:SI_fridge-wiring}. For clarity, connections between the control PC and the various instruments are omitted. 

\section{Control of On-site Energies}
\label{SI:flux_control}

\subsection{Flux tuning and crosstalk calibration}

The lattice experiments require precise and rapid tuning of the on-site frequency (the transmon qubit $\omega_{01}$) for each lattice site. The \textit{dc} flux bias lines are used to statically tune the sites to a target frequency, while the fast flux bias lines are used to provide additional dynamical tuning with nanosecond precision.  

The on-site frequencies $\omega_{01}$ are controlled by currents in the flux bias lines. In general, there is substantial cross-talk between flux-bias lines: the amount of flux enclosed in the SQUID loop of each qubit $\phi_i$ is affected by currents in all flux bias lines $I_j$. To change the on-site frequency of each lattice site independently, we must calibrate the crosstalk between all flux lines and all qubits. We assume that the flux crosstalk is linear in the applied currents, such that:

\begin{equation*}
	d\phi_i = \sum_{j=1}^8 \frac{\partial\phi_i}{\partial I_j} dI_j
\end{equation*}
To obtain the crosstalk matrix $M_{ij} = \partial\phi_i/\partial I_j$, we park $Q_i$ at a frequency $\omega_{i}^0$ on the flux slope where the qubit frequency varies linearly with flux for small changes in flux. We then measure the change in frequency $\partial\omega_{i}/\partial I_j$ as the bias current in each of the flux lines $I_j$ is varied. After dividing out the constant flux slope at this particular qubit frequency $\partial\omega_i/\partial\phi_i |_{\omega_i^0} $, we obtain the $i^{th}$ row of the crosstalk matrix $\partial\phi_i/\partial I_j$. These measurements are repeated for all qubits versus changes in all flux line currents. The individual qubit frequencies for the crosstalk calibration are measured by Ramsey interferometry.

The eigenvectors of the inverted crosstalk matrix $M_{ij}^{-1}$ then provide the linear combinations of flux currents that independently tune each qubit, enabling us to calculate the bias currents necessary to bring all qubits to any desired values of $\phi_i$. Finally, the frequency $\omega_i$ to flux $\phi_i$ conversion for each qubit can be obtained by either fitting the measured qubit spectra versus flux $\omega_i = \omega_i(\phi_i)$ to a Jaynes-Cummings model, or by measuring a linear slope if only tuning the qubits over a small frequency range close to the linear part of the flux slope.

We show the measured \textit{dc} flux crosstalk matrix in Table ~\ref{tab:SI_flux-crosstalk}. Rows 1-7 are on-chip flux lines proximal to $Q_1-Q_7$ respectively (the external \textit{dc} coil tunes all qubits with similar flux to current slopes, not shown here). The columns of the matrix have been been normalized to the diagonal elements to show the relative magnitude of the cross talks. In Table~\ref{tab:SI_flux-crosstalk2}, we show the fast \textit{dc} flux crosstalk matrix for the on-chip flux lines. These values are measured and accurate for fast flux pulses of length as long as $20\,\mu$s, longer than all relevant experimental sequences in this work.

From the measurements, the fast flux crosstalk is significantly smaller than the DC crosstalk. We attribute this to the fact that the on-chip flux lines are individually grounded near each qubit which causes high frequency (fast) flux signals to reflect back into the flux lines and appear more ``localized'' for other qubits, compared to the DC biasing currents which flow into the superconducting ground plane and may trace out peculiar routes as they try to follow a return path of least impedance. 

We did not observe any noticeable non-linearity in the crosstalk for the on-chip flux lines within our measurement accuracy, but they could potentially arise as higher order effects especially at higher bias currents. For future experiments, the cross talk can be greatly reduced by careful engineering of the flux lines (e.g. the Google/UCSB team reported $\lesssim0.3$\% crosstalk in a 9 qubit chain~\cite{Neill2018}), while keeping high flux sensitivity (low bias current). 

We achieve independent qubit frequency tuning with the precision limited by the accuracy of the flux crosstalk inversion and the time-domain pulse shape of the flux-bias (next section). We measure typical discrepancies between the intended on-site frequencies and the measured values of $\delta\omega_{01} \lesssim 2\pi\times 200$\,kHz. In addition, near a degenerate lattice, we can directly measure the normal mode frequencies of the coupled lattice~\cite{ma2017hamiltonian} from which we can back out the exact on-site disorder and compensate with the flux biases. After the compensation, we estimate an residual disorder of $\delta\epsilon = \delta\omega_{01} \lesssim 2\pi\times 100$\,kHz. 

\begin{table}
    \centering
    \begin{minipage}{0.5\textwidth}
        \centering
        {\renewcommand{\arraystretch}{1.4}%
    	\begin{tabularx}{0.8\textwidth}{ |Y||*{7}{Y|} }
    	    \multicolumn{8}{c}{$\partial \phi_i / \partial I_j \hspace{0.1in} (\%)$} \vspace{0.05in} \\  \hline
    		\diagbox[height=0.2in]{i  }{  j} & 1 & 2 & 3 & 4 & 5 & 6 & 7 \\ \hline\hline
    		1 & 100.0 & 17.9 & -1.8 & -33.7 & -41.2 & -37.6 & -27.4    \\ \hline
            2 & 22.4 & 100.0 & 0.8 & -37.9 & -41.9 & -38.6 & -28.6    \\ \hline
            3 & 26.3 & 24.0 & 100.0 & -49.0 & -51.2 & -47.0 & -30.4   \\ \hline
            4 & 17.2 & 16.8 & 9.0 & 100.0 & -44.4 & -37.0 & -22.0   \\ \hline
            5 & 25.5 & 22.9 & 14.4 & -39.9 & 100.0 & -64.4 & -30.9    \\ \hline
            6 & 15.8 & 16.7 & 12.3 & -20.1 & -41.9 & 100.0 & -27.1   \\ \hline
            7 & 9.3 & 14.6 & 17.0 & 12.3 & -15.4 & -37.7 & 100.0    \\ \hline
    	\end{tabularx}
    	}
        \caption{Measured static \textit{dc} flux crosstalk matrix.}
    	\label{tab:SI_flux-crosstalk}
    \end{minipage}%
    \begin{minipage}{0.5\textwidth}
        \centering
        {\renewcommand{\arraystretch}{1.4}%
    	\begin{tabularx}{0.8\textwidth}{ |Y||*{7}{Y|} }
    	    \multicolumn{8}{c}{$\partial \phi_i / \partial I_j \hspace{0.1in} (\%)$} \vspace{0.05in}  \\ \hline
    		\diagbox[height=0.2in]{i  }{  j} & 1 & 2 & 3 & 4 & 5 & 6 & 7 \\ \hline\hline
    		1 & 100.0 & -2.1 & -5.8 & -3.5 & -2.9 & -2.6 & -0.9 \\ \hline
            2 & 4.5 & 100.0 & -5.1 & -3.9 & -3.2 & -3.0 & -0.8 \\ \hline
            3 & 4.4 & 4.6 & 100.0 & -6.1 & -4.2 & -3.4 & -0.9 \\ \hline
            4 & 6.3 & 6.0 & 6.7 & 100.0 & -16.3 & -8.2 & -2.5 \\ \hline
            5 & 4.5 & 4.0 & 5.7 & -2.5 & 100.0 & -8.4 & -1.4 \\ \hline
            6 & 3.1 & 3.0 & 5.6 & -0.8 & -2.3 & 100.0 & -2.8 \\ \hline
            7 & 2.6 & 3.5 & 9.0 & 5.5 & -0.8 & -7.4 & 100.0 \\ 
    		\hline
    	\end{tabularx}
    	}
        \caption{Measured fast \textit{dc} flux crosstalk matrix.}
        \label{tab:SI_flux-crosstalk2}
    \end{minipage}
\end{table}

\subsection{Time domain flux pulse shaping}

\subsubsection{Correcting flux pulse response}

For dynamical tuning using the fast flux bias, we would like the individual sites to see precise flux pulse shapes (e.g. a step function detuning with finite ramp). The fast flux pulses are generated with a $600$\,MS/s AWG (PXDAC4800), and the various elements in the flux line circuit between the AWG and the transmon qubit lead to significant pulse distortion. These elements include room temperature filters and electronics; cryogenic lines, attenuators and filters; and the on-chip flux line that carries the flux bias signal from the RF connector to the vicinity of the transmon. For practical purposes, these distortions are linear in the flux pulse amplitude. So by measuring the transfer function of the linear response of the flux circuit, we can construct a deconvolution kernel applied to the AWG pulses to compensate and correct for the flux pulse shape. We follow a procedure similar to that described in Ref.~\cite{Johnson2011thesis}. 

In our case, the main contribution of the flux pulse distortion comes from the effective low-pass effect of the filters and RF lines, and the contribution from the AWG ouput's intrinsic distortion is relatively negligible. We measure the distortion directly with time-resolved qubit spectroscopy: We apply a step flux pulse to the AWG output, and measure the response of the qubit frequency to the step pulse as a function of time. The qubit charge excitation pulse has a Gaussian shape with $\sigma=20$\,ns truncated at $\pm 2\sigma$, and weak enough to only excite a fraction of the qubit. By fitting the qubit spectrum at each time with a Lorentian, we obtain the instantaneous qubit frequency. For a small flux step applied with the qubit frequency on a linear slope, the time resolved qubit frequency trace gives the filtered step response seen by the qubit. The time resolution of the spectroscopy is limited by the length of the excitation pulse to tens of ns.

The measured time-domain response is Fourier transformed to get the frequency-domain response. Because of the finite resolution of the qubit spectroscopy, we keep only the low frequency response up to a cutoff $f_c \sim 25$\,MHz, which is then inverted and Fourier transformed back to obtain the time-domain kernel for pulse compensation. To achieve compensated response at the qubit, the AWG output is then set to the target pulse shape convoluted with the time-domain kernel. The cutoff $f_c$ insures that the high frequency response remains unaltered.

\begin{figure}
	\includegraphics[width=0.45\columnwidth]{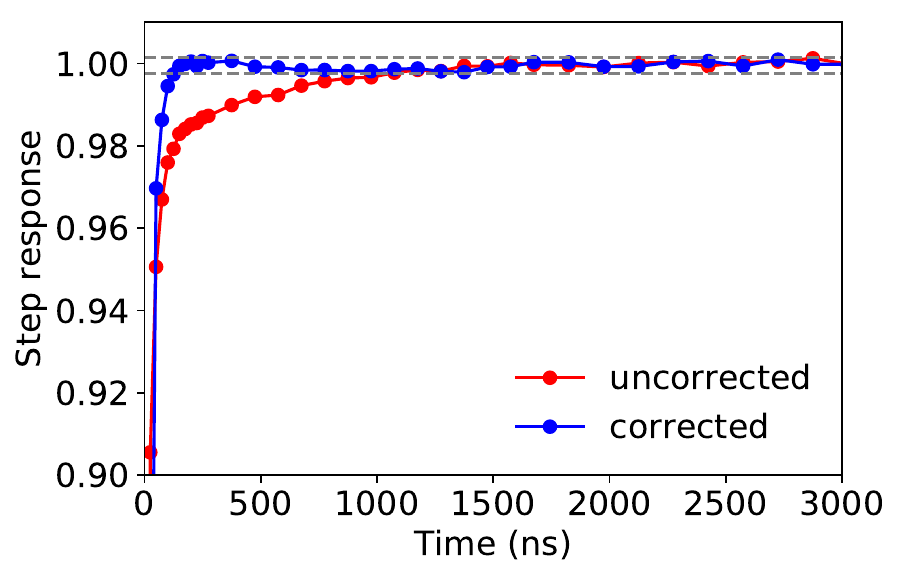}
	\caption{\textbf{Time domain pulse correction for fast flux biasing.} The qubit frequency response after an applied step to the fast flux bias is measured using time resolved qubit spectroscopy. Shown are results before (red) and after (blue) pulse compensation.
	}
	\label{fig:SI_flux_shaping}
\end{figure}

In Fig.~\ref{fig:SI_flux_shaping}, we show an example of the pulse compensation. The measured step response after compensation settles to within $< 0.5\%$ in less than $100$\,ns.

\subsubsection{Flux balancing pulse}

Experimentally it's been observed that the flux bias lines have residual slow response to fast flux pulses at millisecond or longer time scales, longer than the duration of an experiment cycle. Thus any non-zero net flux current applied during the experiment sequence could lead to unwanted drifts of the qubit frequency as the experiment is continuously run. To avoid such effects, we apply a flux balancing pulse at the end of each experimental run such that the net flux currents applied to each flux line always remain zero.

\subsubsection{Estimate short-time flux response}
\label{supp-flux_shping-shorttime}

As discussed above, the procedure for fast flux pulse correction does not reveal information about the short-time response of the flux lines, nor does the compensation we apply alter that short- time response. To estimate the initial ramp speed after applying a step to the flux line, we use a Ramsey-type ``qubit oscilloscope''~\cite{Ciorciaro2017thesis}.

We detect the qubit frequency response to a short square flux pulse (uncompensated), by placing the flux pulse in between the two $\pi/2$ pulses of a Ramsey sequence and measure the extra accumulated qubit phase from the qubit detuning. The delay between the $\pi/2$ pulses are kept long enough so that the qubit has time to return to the initial frequency. The accumulated phase is measured as a function of the length of the square flux pulse, by fitting the phase of the Ramsey fringes. 

\begin{figure}
	\includegraphics[width=0.95\columnwidth]{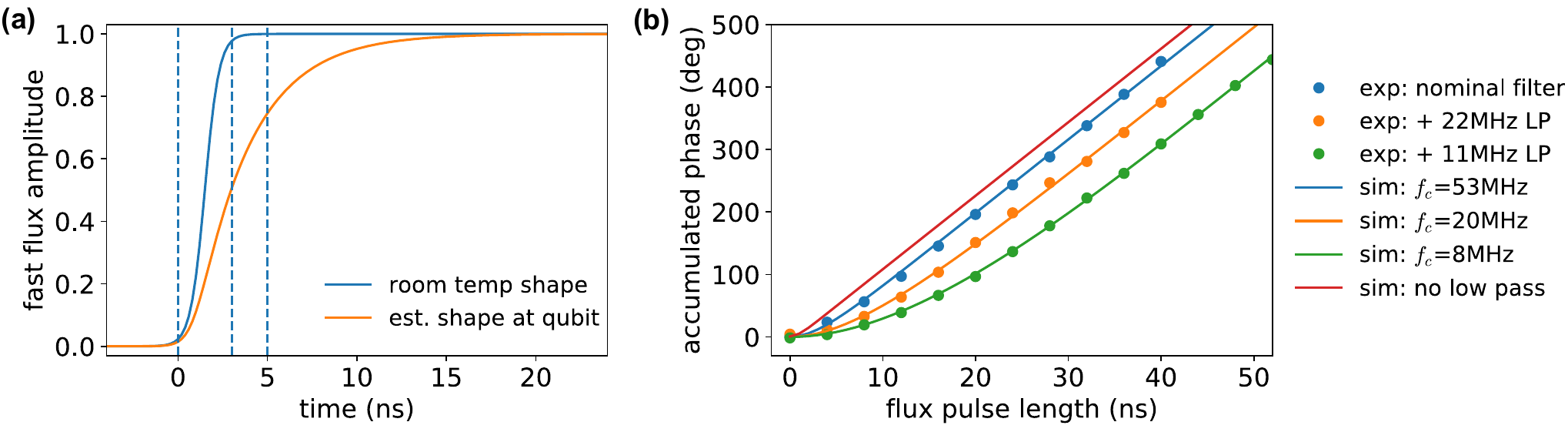}
	\caption{\textbf{Short time response of the fast flux.} \textbf{(a)} The fast flux pulse response to a step at room temperature before entering the fridge (blue) and at the qubit (orange) assuming a simple low pass from the cryogenic components with $f_c$ measured in (b). \textbf{(b)} Accumulated qubit phase as a function of the flux pulse length, measured in a Ramsey type experiment. The bent at short times is a result of the finite ramps of the flux response. From the measured data (blue) we extract the estimated low pass $f_c$. We also plot measurement and simulation when we apply additional low pass filtering at room temperature, showing good agreement.
	}
	\label{fig:SI_ramsey-scope}
\end{figure}

If the qubit detuning is linear with the flux pulse amplitude, then the accumulated phase will be linear in the flux pulse length regardless of the shape of the flux circuit response (because the Ramsey pulses enclose both the rising and falling edges of the pulse, and the filtered response at the two edges are identical with opposite signs). Therefore to obtain information on the short-time flux response, the qubit detuning has to be a non-linear function of the applied flux.
We start the qubit at the flux insensitive lower sweet spot, so that the qubit detuning is quadratic with the applied flux amplitude. We are mostly interested in the initial ramp speed, and the ramps at both edges contribute significantly to the accumulated phase so we cannot simply take the derivative of the accumulated phase to get the frequency response like in Ref.~\cite{Ciorciaro2017thesis}. Instead we directly compare the measured accumulated phase to numerical simulations. The square fast flux pulse from the AWG after all room temperature filtering has a smooth ramp of 3\,ns, that we can fit well with a logistic function shown in Fig.~\ref{fig:SI_ramsey-scope}\textbf{(a)}. We assume that to lowest order, the additional response of the cryogenic lines and filtering is a simple low pass filter with cutoff $f_c$ (and corresponding characteristic time $\tau$). We numerically integrate the expected accumulated phase for different $f_c$, and compare with measurement to extract an estimated $f_c \approx 53$\,MHz ($\tau \sim 3$\,ns) for our fast flux bias lines (Fig.~\ref{fig:SI_ramsey-scope}\textbf{(b)}). The validity of the method is checked by intentionally adding stronger room temperature low pass filters and observe good agreement with measurement and simulation.

This estimated fast flux ramp shape at the qubit (shown in Fig.~\ref{fig:SI_ramsey-scope}\textbf{(a)}) is used in Sec.~\ref{SI:readout+errorbar} below to estimate the readout errors from Landau-Zener crossing between neighboring sites as a result of the finite ramp speed.

\begin{figure}
	\includegraphics[width=0.85\columnwidth]{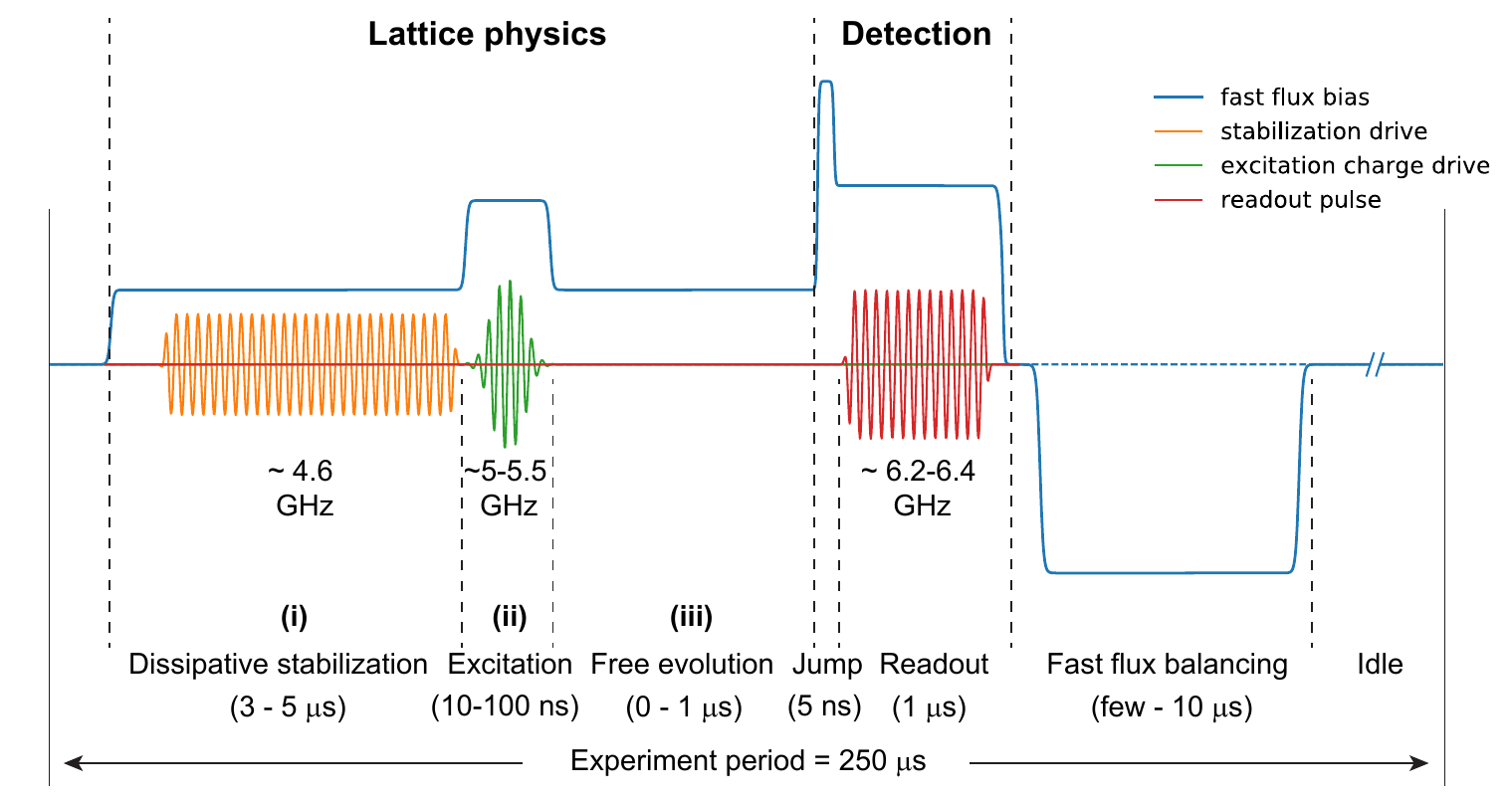}
	\caption{\textbf{Experimental sequence.} We plot the timing of the various drive and flux control pulses for a typical experiment.
	}
	\label{fig:SI_exp-seq}
\end{figure}

\section{Experiment Sequence}
\label{SI:exp_sequence}

We briefly describe the experimental sequence for a typical experiment: first the quantum many body state of interest is prepared and time-evolved in the lattice (``lattice physics'');  then the the state of the lattice is measured (``detection''). 

At the beginning of each experiment, all microwave signals are idle and the lattice is empty. Constant \textit{dc} flux biases tune the sites to near the lattice frequency, while the fast flux biases are applied additionally during the sequences to achieve dynamical tuning of all sites.

To perform different lattice experiments, segments with different lattice detunings and drive pulses are concatenated together for the ``lattice physics.''. In Fig.~\ref{fig:SI_exp-seq} we illustrate a generic sequence: (i) first a stabilization drive is used to dissipatively prepare a many-body state across the degenerate lattice; (ii) then one site is quickly (much faster than $1/J$) detuned from it's neighbors and charge drive pulses are applied to put an individual excitation into the spectrally isolated site; (iii) the detuned site is then brought back into resonance with the rest of the lattice and we observe the dynamics of the many-body state by varying the time of the free evolution before the readout. During step (ii), we typically detune all sites from their neighbors in order to ``freeze'' all the tunneling dynamics; more than one excitations can be created either sequentially, or simultaneously by frequency multiplexed drive pulses via the common charge-drive line.

At the end of the lattice evolution, all drive pulses are turned off, and the fast flux biases are ``jumped'' rapidly (with maximum amplitude change and maximum bandwidth to achieve fastest ramp rate) to detune the measured lattice site away from its neighbors, effectively freezing its on-site occupancy. The readout microwave pulse is then applied for a duration of $1\,\mu$s during which the output from the device is acquired using the heterodyne setup. After the readout, the flux balancing pulses are applied with varying length (few $\mu$s) to null the net current flowing into each flux line during the sequence. The details and error estimation of the lattice readout are described in the next section.

The detuning ramps used during the ``lattice physics'' are typically $10-20$\,ns in duration, linear in the (uncorrected) fast flux amplitude but smooth at the qubits from the effective low pass of the flux lines. All charge excitation pulses are Gaussians truncated at $\pm 2\sigma$, while the stabilization drives are square pulses with few ns soft ramp-up/down of the amplitude.

The experiment is then repeated with a cycle period of $250\,\mu$s, leaving enough idle time after each sequence for the lattice sites to decay back to their (thermal equilibrium) ground states. This corresponds to a repetition rate of 4\,kHz. For figures in the main text, each single data point (averaged over 5000-8000 runs) takes about 1-2 seconds, while a full 2D plot takes a few hours.

\begin{figure}
	\includegraphics[width=0.6\columnwidth]{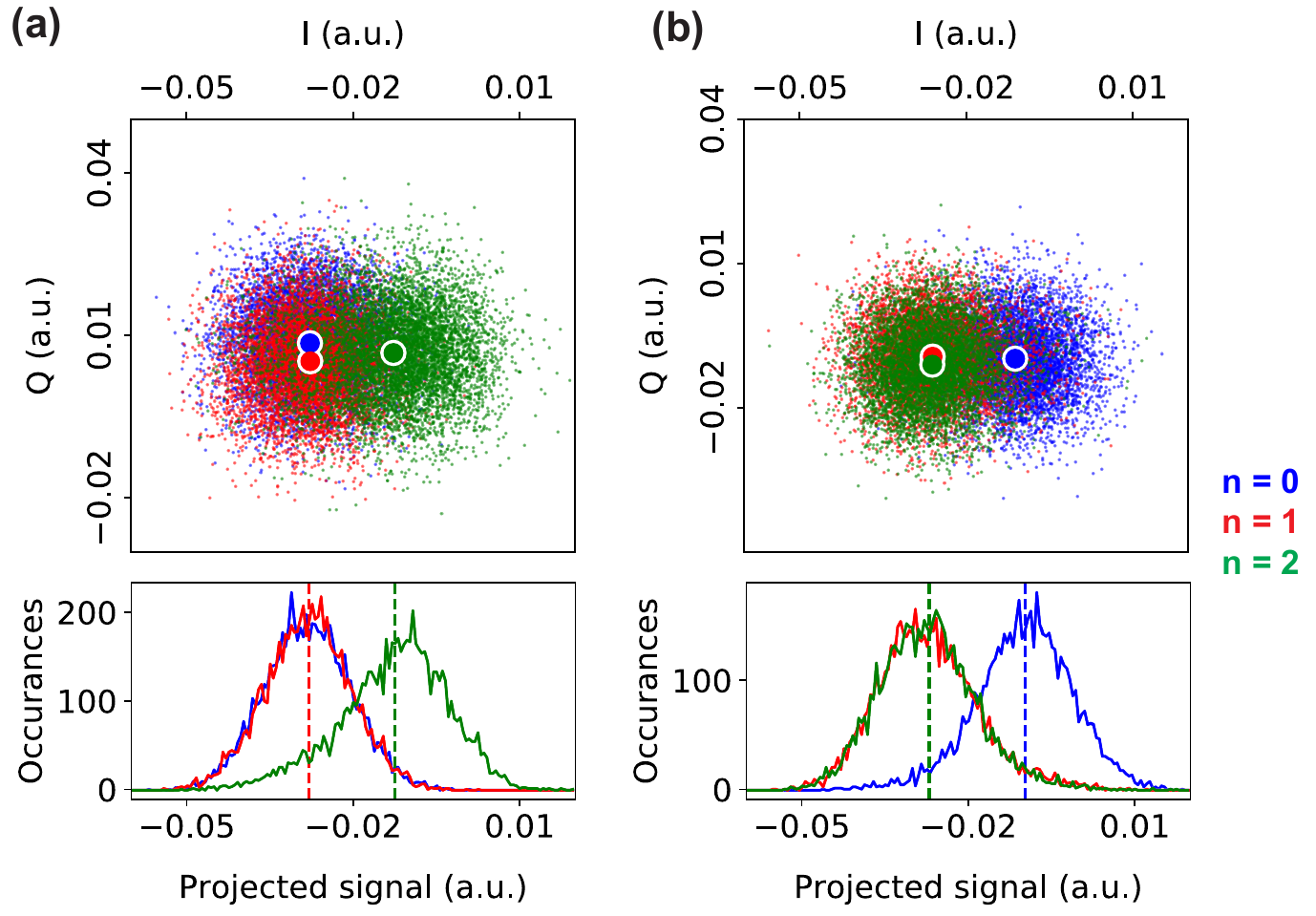}
	\caption{\textbf{Qubit readout.} The digitally-homodyned quadrature signals for the dispersive readout of site $Q_3$. Each of the $6000$ data point in the scatter plot corresponds to the time averaged signal of a single shot readout with the qubit initialized in one of states $n=0,1,2$. We choose readout frequencies $\omega_{\textrm{read}}^f = 2\pi\times 6.43145$\,GHz \textbf{(a)} and $\omega_{\textrm{read}}^g = 2\pi\times 6.43355$\,GHz \textbf{(b)} to distinguish only state $\ket{0}$ or $\ket{2}$. Dots in the scatter plots indicate average signal of each state. The global phases have been adjusted so that the desired projection is onto the horizontal axis. On the bottom we show histograms of the projected signal, where now the average signals ($\upsilon$s and $\nu$s) are indicated by the dotted lines.
	}
	\label{fig:SI_readout}
\end{figure}

\section{Readout of the Bose-Hubbard Lattice}
\label{SI:readout+errorbar}

The on-site occupancy of the prepared many-body state is obtained by measuring the qubit state dependent dispersive shift of the readout resonator, after a rapid detuning (``jump'') of the measured site to isolate it from the rest of the lattice. We calculate the expectation values of $P_n$ from the averaged readout signal, calibrated with separately prepared qubit states $\ket{n}$. We consider two contributions to readout errors: (1) the ``jump'' during which Landau-Zener crossings between the measured site and its neighbor(s) lead to population transfer between neighbors, and (2) the dispersive readout where errors in the calibration states due to imperfect $\pi$-pulses used to prepare them lead to errors when mapping the averaged readout signal to on-site occupancies.

\subsection{Heterodyne dispersive readout}

The on-site occupancy (i.e. transmon state) is measured with the state-dependent dispersive shift of the readout resonator. The bare frequencies of the individual readout resonators $\omega_{\mathrm{read}}$, their linewidths $\kappa_{\mathrm{read}}$ and coupling to the qubits $g_{\mathrm{read}}$ are listed in Table.~\ref{tab:SI_coherences}. During readout, the measured site is typically detuned to $\delta \sim 1$\,GHz below its readout resonator. In this strongly dispersive regime where the differential dispersive shift between successive qubit states $2\chi \approx 2\pi\times 0.8$\,MHz exceeds the $\kappa_{\mathrm{read}} \sim 2\pi\times 0.4$\,MHz, we can distinguish different occupation states of the qubit by measuring the complex reflection off of the readout resonator at several probe frequencies.

We perform the readout at relatively weak probe powers ($<10$ intra-resonator photons) for a duration of $1\,\mu$s. The heterodyne signal recorded by the acquisition ADC card is converted to quadrature I/Q signals via digital homodyning and averaged over the readout window. The  distinguishability of different transmon states from a single-shot measurement is limited to $\sim 40-50\%$ by thermal noise of the HEMT amplifier. Therefore we use averaged readout signals (of typically $\sim 5000$ runs of the experiment) to extract the expectation values of the on-site occupancies. The scales for $P_n$ are calibrated by measuring $\ket{0},\ket{1},\ket{2}$ states (transmon $\ket{g},\ket{e},\ket{f}$ states) prepared using resonant $\pi$-pulses (see section below on effect of thermal population). Introduction of a near quantum limited amplifier for the output, e.g. by using a traveling wave parametric amplifier (TWPA) will permit high-fidelity single-shot readout~\cite{Macklin2015}.

For each qubit, we choose two different probe frequencies $\omega_{\textrm{read}}^g$ and $\omega_{\textrm{read}}^f$ such that for $\omega_{\textrm{read}}^g$ we achieve optimal distinguishability between $\ket{0}$ and $\ket{1}/\ket{2}$ while minimizing the distinguishability between $\ket{1}$ and $\ket{2}$. Similarly $\omega_{\textrm{read}}^f$ is chosen to distinguish $\ket{2}$ from $\ket{0}/\ket{1}$. The measured I/Q signal is then projected onto an axis in I/Q space that minimizes the distinguishability between $\ket{1}$ and $\ket{2}$ ($\ket{0}$ and $\ket{1}$) for $\omega_{\textrm{read}}^g$ ($\omega_{\textrm{read}}^f$). This allows us to measure the population in $\ket{0}$ and $\ket{2}$ with $\omega_{\textrm{read}}^g$ and $\omega_{\textrm{read}}^f$, respectively. We denote the projected signals by $\upsilon$ and $\nu$ when probing at $\omega_{\textrm{read}}^g$ and $\omega_{\textrm{read}}^f$, respectively.
This process is illustrated in Fig.~\ref{fig:SI_readout}, where we plot the readout signal from one of the qubits, corresponding to a particular lattice site.

For reading out the lattice, we measure one qubit at a time, and separately for $\omega_{\textrm{read}}^g$ and $\omega_{\textrm{read}}^f$ each repeated $\sim 5000$ times to obtain the averaged signals. During readout the measured qubit is spectrally isolated with neighbors far detuned, allowing for clean $\pi$-pulses used for initializing the states for readout calibration.

\subsection{Extraction of on-site occupancy }

The finite effective temperature of the qubits means that the states prepared with resonant $\pi$-pulses are not perfect Fock states $\ket{n}$, but statistical mixtures originating from the initial thermal population $n_{\mathrm{th}}$. Therefore we need to take into account $n_{\mathrm{th}}$ when using the average readout signal from the prepared calibration states to extract the on-site occupancies. Given our qubit frequencies and low effective qubit temperatures, it is a good approximation that all thermal excitation promotes qubits to the $n=1$ state, with negligible thermal excitation to $n>1$. Aside from the thermal ground state, we prepare $3$ more states by applying resonant $\pi-$pulses, with density matrices given by:
\begin{alignat*}{5}
    & \text{Thermal ground state:} \quad & A & \enskip=\enskip & (1-n_{\mathrm{th}}) & \ket{g}\bra{g} \enskip+\enskip & n_{\mathrm{th}} & \ket{e}\bra{e} \enskip+\enskip & 0 \ket{f}\bra{f}\\
    & \text{A} + \pi_{01} \text{ pulse:} \quad & B & \enskip=\enskip & n_{\mathrm{th}} & \ket{g}\bra{g} \enskip+\enskip & (1-n_{\mathrm{th}}) & \ket{e}\bra{e}  \enskip+\enskip & 0 \ket{f}\bra{f}\\
    & \text{A} + \pi_{01} + \pi_{12} \text{ pulses:} \quad & C & \enskip=\enskip & n_{\mathrm{th}} & \ket{g}\bra{g} \enskip+\enskip & 0 & \ket{e}\bra{e} \enskip+\enskip &  (1-n_{\mathrm{th}}) \ket{f}\bra{f}\\
    & \text{A} + \pi_{12} \text{ pulse:} \quad & D & \enskip=\enskip & (1-n_{\mathrm{th}}) & \ket{g}\bra{g} \enskip+\enskip & 0 & \ket{e}\bra{e} \enskip+\enskip &  n_{\mathrm{th}} \ket{f}\bra{f}
\end{alignat*}

If the projected heterodyne readout voltages for the states are $\upsilon_i$ and $\nu_i$ ($i \in A,B,C,D$) when probing at $\omega_{\textrm{read}}^g$ and $\omega_{\textrm{read}}^f$ respectively, we see that:
\begin{equation*}
    \frac{n_{\mathrm{th}}}{1-n_{\mathrm{th}}} = \frac{\nu_D-\nu_A}{\nu_C-\nu_A}
\end{equation*}
from which we obtain $n_{\rm th}$. Then using the measurements of the calibration states, we can extract the on-site occupancy of an arbitrary state $\psi$, with measured readout signals $\upsilon_\psi$ and $\nu_\psi$. By defining the uncalibrated populations $p_0(\psi) = (\upsilon_\psi - \upsilon_B)/(\upsilon_A - \upsilon_B)$ and $p_2(\psi) = (\nu_\psi - \nu_A)/(\nu_C - \nu_A)$, we get the on-site occupancies $P_n$ assuming all population resides in $n\leq 2$: 
\begin{align*}
    P_0(\psi) &= p_0(\psi) (1-2 n_{\mathrm{th}}) + n_{\mathrm{th}}\\
    P_2(\psi) &= p_2(\psi) ( 1 - n_{\mathrm{th}}) \\
    P_1(\psi) &= 1 - P_0(\psi) - P_2(\psi)
\end{align*}

Given the calibration $\pi$-pulse errors of $\sim 0.5\%$, we estimate a $\pm 1\%$ uncertainty on the extracted $P_1$ from readout calibrations. 

\subsection{Landau-Zener transfer during measurement}

The rapid ``jump'' of the qubit frequency before the readout detunes the measured qubit far away from its neighbors ($|\delta| > |U|$), not only to freeze tunneling dynamics, but also to suppress neighbor-occupancy-dependent shifts of the measured qubit's levels (which can in turn shift the averaged readout signals). During the ``jump'' ramp, there will be Landau-Zener population transfer whenever the measured qubit's levels cross any of its neighbors' levels, which leads to discrepancies between the measured on-site occupancy and the occupancy of the original lattice prior to the jump. In our cases, we typically jump the measured qubit up in frequency by $\delta > 2|U|$ relative to its neighbors. Thus the on-site population is affected primarily by two Landau-Zener crossings: initially when $\omega_{01}^Q = \omega_{01}^{\textrm{NN}}$ (corresponding to half a LZ crossing); and midway through the jump when $\omega_{12}^Q = \omega_{01}^{\textrm{NN}}$. Here $Q$ is the measured site, and $NN$ any of its neighbors.

We can estimate these transfer probabilities by numerically solving the dynamics of the measured qubit and its neighbors, using the known lattice parameters and detunings used in the experiments and the estimated short-time form of the jump (Sec.~\ref{supp-flux_shping-shorttime}). The magnitudes of these population transfers are dependent on the occupancy of the measured qubit as well as those of its neighbors. For our experimental parameters, the worst cases happen at higher lattice fillings with Bose-enhanced tunneling (in our case $n=2$), and when neighbors with occupancies that differ by one come into resonance during the jump. The induced population change on the measured site can be as much as $10-20\%$ in such cases, leading to significant errors when measuring states with large number fluctuations (i.e. a superfluid). However for the single stabilized site or the $n=1$ Mott phase we study in this work, these Landau-Zener transfers can be much less-- on the few percent level. Therefore it is possible to use the measured on-site populations plus the numerics to estimate the likely (rather than worst-case)  population transfer that occurred during the detuning jumps. Here we assume no coherence between lattice sites, which is reasonable given that the stabilization infidelities are primarily thermal excitations from the reservoir.

\medskip
\textbf{Readout error estimates for data:}

\textbf{(1)} For single site stabilization with the ``one transmon'' scheme, the stabilized site ($Q_1$) is jumped up in frequency by $\sim 700$\,MHz for readout. If we take on-site states of $\rho_S = 0.09\,|0\rangle\langle0| + 0.81\,|1\rangle\langle1| + 0.10\,|2\rangle\langle2|$ from the measurement (as a statistical mixture of the different fock states) and $\rho_R = 0.925\,|0\rangle\langle0| + 0.075\,|1\rangle\langle1|$ (the steady state reservoir due to thermal population), the numerically calculated on-site state after the jump reads $\rho_S^\prime = 0.093\,|0\rangle\langle0| + 0.811\,|1\rangle\langle1| + 0.095\,|2\rangle\langle2|$. Thus in this case, we estimate that the Landau-Zener crossings during the jump change the occupancies by $<0.5\%$ ($0.1\%$ on $P_1$). The measured optimal fidelity after the jump is $P_1 = 0.81$ ($P_0 = 0.09$, $P_2 = 0.10$). With readout uncertainty of $\approx 1\%$ from pulse calibration discussed previously, we therefore place an estimated total errorbar on the optimal ``one trasmon'' scheme fidelity as $P_1 = 0.81(\pm 0.01)$. Here the population transfer is small because the stabilized site's $0-1$ transition starts already detuned, while the $n=2$ transition and the resonant reservoir are rarely populated. 

\textbf{(2)} In the ``two transmon'' scheme, again using actual experimental parameters and the estimated fast flux shape to numerically calculate the most likely Landau-Zener transfers, we find that near the steady state optimal fidelity, the stabilized site $P_0$ should change by $\leq 1\%$; $P_1$ will be reduced by up to $5\%$ that mostly turn into $P_2$. Changes in $P_0$ is significantly less because in the experiment the $0-1$ transition of the stabilized site remains detuned throughout the jump, and therefore also less sensitive to exact details of the ramp. This allows us to use the measured $(1-P_0)$ as an upper bound for the fidelity. The measured optimal fidelity (after the jump) is $P_1 = 0.89$ ($P_0 = 0.08$, $P_2 = 0.03$). Therefore we place a total estimated errorbar as $P_1 = 0.89 (+0.04/\text{\textminus}0.01)$.

\textbf{(3)} For the dissipatively stabilized Mott fidelity, the numerically estimated population transfers are similar to the ``two transmon'' scheme above. On average over all sites of the Mott, changes in $P_0$ remains small ($<1\%$) while $P_1$ ($P_2$) is expected to reduce (increase) by a few percent. Using the measured $\langle 1-P_0 \rangle = 0.90(\pm 0.01)$ as an upper bound, this puts the errorbar for the optimal average Mott fidelity at $\langle P_1\rangle = 0.88(+0.03/\text{\textminus}0.01)$. The measured $\langle P_2\rangle= 0.02(+0.01/\text{\textminus}0.02)$ indicates that infidelities in the Mott state are predominantly holes.

\medskip
\textbf{Future improvements:}

Moving forward, it will be essential to explore states with stronger number fluctuations, where $J/U$ is larger. To ensure that this does not induce large readout errors, we could take several approaches (possibly a combination of them): (1) tunable couplers~\cite{lu2017universal,roushan2017spectral}, which allow the tunneling to be turned off completely during the detuning sweeps and the readout, eliminating the population transfers. (2) RF flux modulation on the measured qubit to create an instantaneously detuning of the qubit (modulate at frequency difference of lattice location and the readout location) without the need to ramp through the intermediate Landau-Zener crossings. (3) faster flux biasing, to be able to detune the lattice sites more rapidly and reduce any residual Landau Zener transfers. 

With the current device and detuning method, improvement is possible by using near-quantum limited amplifiers~\cite{Macklin2015} to achieve high-fidelity single shot readout, so that we can detune the measured qubit up in frequency by only $\sim |U|/2$ during readout. This avoids the crossing between the $n=2$ of the measured site and $n=1$ of its neighbor, which is where the majority of the population transfer happens. Without single shot readout, at a relatively small neighbor detuning of $U/2$ the measured qubit state frequencies are dependent on its neighbors' state which changes the average readout signal significantly. Thus the calibration signals measured when the neighbors are all empty (what we do currently) can not be used to properly extract the on-site occupancies of a generic state in the lattice.

\section{Qubit and Bose-Hubbard Lattice Characterization}
\label{SI:qubit_&_BH_cal}

\subsection{Qubit coherences}
\label{SI:sub-coherences}

In Table.~\ref{tab:SI_coherences} we list measured relaxation times $T_1$, dephasing times $T_2^*$, and the corresponding decoherence rates for all lattice sites (transmon qubits). Sites $Q_2$ to $Q_8$ are measured near the nominal lattice frequency of $\sim 4.7$\,GHz; while $Q_1$ is measured at $\sim 5.3$\,GHz to avoid being Purcell limited by the reservoir linewidth. For $T_1$, values in parenthesis indicate standard deviations of 10 measurements taken over one day. The dephasing times $T_2^*$ are limited by flux noise from the external flux bias sources, especially the AWGs that produce the fast flux pulses. We measure $T_2^* \approx 2-4\,\mu$s for $Q_1 - Q_7$, depending on the output amplitude of the these AWGs. $Q_8$ does not have a direct on-chip flux line, thus has a longer $T_2^*$. We expect quieter biasing electronics and additional filtering to reduce the dephasing rates in our device.

\begin{table}
	{\renewcommand{\arraystretch}{1.4}
	\begin{tabularx}{0.6 \textwidth}{| Y  Y | Y | Y | Y | Y | Y | Y | Y | Y |} 
      \hline
       ~ & ~ & $Q_1$ & $Q_2$ & $Q_3$ & $Q_4$ & $Q_5$ & $Q_6$ & $Q_7$ & $Q_8$    \\ \hline\hline
       \multicolumn{2}{|c|}{$T_1$ ($\mu$s)} & 22(4)& 19(4) & 30(3) & 40(3) & 34(4) & 42(3) & 19(3) & 36(5)  \\ \hline
       \multicolumn{2}{|c|}{$\Gamma_1/2\pi$ (kHz)} & 7.2 & 8.4 & 5.3 & 4.0 & 4.7 & 3.8 & 8.4 & 4.4       \\ \hline
      \multicolumn{2}{|c|}{$T_2^*$ ($\mu$s)} & 2 - 4 & 2 - 4 & 2 - 4 & 2 - 4 & 2 - 4 & 2 - 4 & 2 - 4 & 5 \\ \hline
     \multicolumn{2}{|c|}{$\Gamma_\phi^*/2\pi$ (kHz)} & 40-80 & 40-80 & 40-80 & 40-80 & 40-80 & 40-80 & 40-80 & 30  \\ \hline

      \multicolumn{2}{|c|}{$U/2\pi$ (MHz)} & -254.3 & -258.6 & -254.1 & -160.0 & -253.2 & -247.7 & -252.0 & -252.4  \\ \hline
      \multicolumn{2}{|c|}{$J_{i-1,i}/2\pi$ (MHz)} & 16.30 & 12.68 & 6.34 & 6.47  & 6.18  &  6.33  & 6.37  & 6.09 \\ \hline\hline
      \multicolumn{2}{|c|}{$n_{\textrm{th}}$ } & 0.07 & 0.06 & 0.03 & 0.05  & 0.04  &  0.06  & 0.02  & 0.06 \\ \hline\hline
\multicolumn{2}{|c|}{$\omega_{\textrm{read}}/2\pi$ (GHz)} & 6.474 & 6.367 & 6.467 & 6.346  & 6.430  &  6.310  & 6.381  & 6.261 \\ \hline
      \multicolumn{2}{|c|}{$g_{\textrm{read}}/2\pi$ (MHz)} & 70 & 69 & 70 & 66  & 70  &  70  & 70  & 68 \\ \hline
      \multicolumn{2}{|c|}{$\kappa_{\textrm{read}}/2\pi$ (MHz)} & 0.50 & 0.40 & 0.44 & 0.43  & 0.40  &  0.44  & 0.42  & 0.33 \\ \hline
    \end{tabularx}
    }
\caption{\textbf{Measured lattice parameters.} Here we list all lattice site coherences, Bose-Hubbard parameters, steady state thermal populations, and readout parameters.}
\label{tab:SI_coherences}
\end{table}

\subsection{Bose-Hubbard parameters}
\label{SI:sub-BH_params}
\subsubsection{Tunneling rates $J$}
\label{SI:J-cal}

To measure the nearest neighbor tunneling matrix element $J_{ij}$ we observe the dynamics of a single photon in an isolated double-well potential formed by the two neighboring lattice sites $i$ and $j$. Starting with an empty lattice, we excite one photon into site $i$ by applying a $\pi$-pulse at $\omega_{01}^i$ while all other lattice sites are far detuned. We then rapidly tune site $j$ into resonance with site $i$ and measure the population in each site as a function of the evolution time. The population oscillation should have a frequency of $2J_{ij}$. In Fig.~\ref{fig:supp_JU_cal}\textbf{(a)}, we show one such measured oscillation for $J_{56}$. All tunneling rates are listed in Table.~\ref{tab:SI_coherences}. The value for $J_{12}$ was designed to give better stabilizer performance. The tunneling $J_{R1}$ between $Q_1$ and the reservoir is measured by tuning $Q_1$'s frequency across the reservoir lossy resonator, and observing the avoided crossing (of splitting $2J_{R1}$) in the reflection spectra of the lossy resonator. Next nearest neighbor tunnelings are suppressed by factors of $\sim 7-10$, based on finite element simulations of the microwave circuit.

\subsubsection{On-site interactions $U$}

\begin{figure}
    \includegraphics[width=0.9\textwidth]{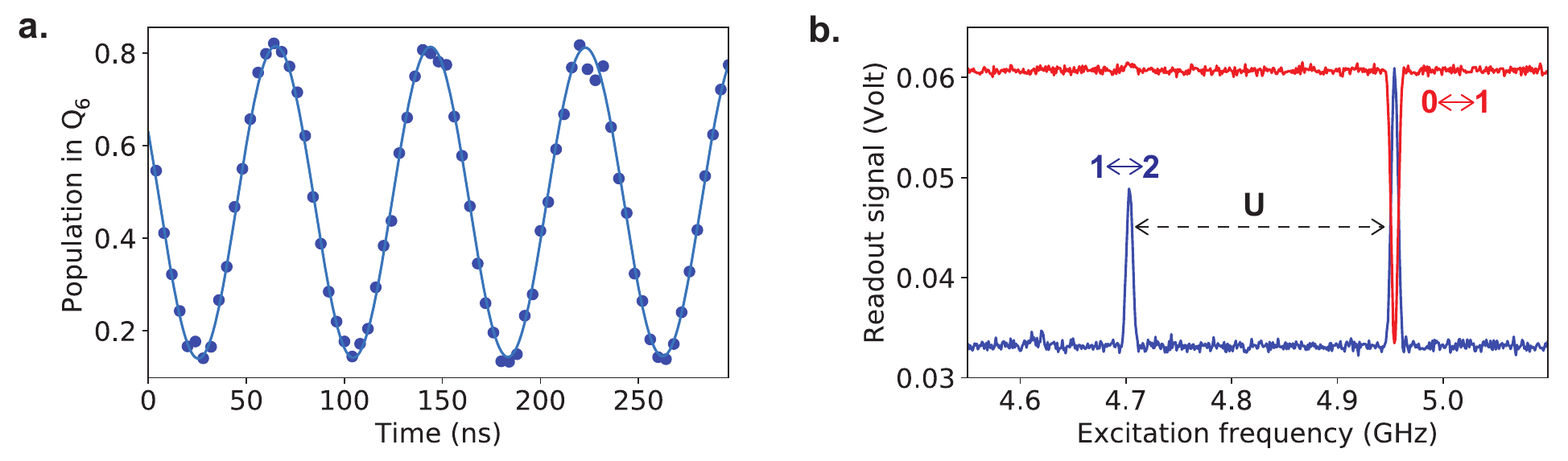}
    \caption{\textbf{Characterization of the Bose-Hubbard parameters.} \textbf{(a)} Oscillation of a single photon in a resonant double well of $Q_5$ and $Q_6$. The fitted oscillation frequency gives $2J_{56}$. The reduced contrast and the offset at $t=0$ are results of the finite speed ramps of the on-site energies. \textbf{(b)} Excitation spectra of an isolated lattice site, with the site initially empty (red) or filled with one photon (blue). The difference in frequency between the two observed transitions ($\omega_{01}$ and $\omega_{12}$) reveals the effective on-site interaction $U$.} 
    \label{fig:supp_JU_cal} 
\end{figure}

The effective on-site photon-photon interaction energy is $U=\omega_{12}-\omega_{01}$, given by the anharmonicity of the qubits that make up the lattice sites. For transmon qubits, the anharmonicity is negative, corresponding to the realization of an attractive Bose-Hubbard lattice. The qubit transition frequencies and thus $U$ can be measured precisely from Ramsey experiments. The transitions can also be probed in the excitation spectra of a single spectrally isolated lattice site, as shown in Fig.~\ref{fig:supp_JU_cal}\textbf{(b)}. Starting with an initially empty site, we drive an excitation pulse at varying frequencies and measure the response of the readout cavity (blue trace); here we observe for $Q_8$ the transition at $\omega_{01} = 2\pi\times 4.955$\,GHz. The excitation pulse is a truncated Gaussian that drives a $\pi$-pulse on the $\omega_{01}$ resonance. The width of the peak is Fourier limited by the pulse spectral width. To probe the $1\rightarrow 2$ transition, we first drive a $\pi$-pulse on the  $0\rightarrow 1$ transition (at $\omega_{01}$) followed by a second excitation pulse with varying frequency; the observed new transition is located at $\omega_{12} \approx 2\pi\times 4.704$\,GHz, indicating an effective on-site interaction $U = - 2\pi\times 251$\,MHz.

The on-site interaction $U$ at each site is listed in Table.~\ref{tab:SI_coherences}, measured at the nominal lattice frequency of $\omega_{01} \sim 2\pi\times 4.7$\,GHz. The anharmonicity of transmon qubits changes slightly as the qubit frequency is tuned~\cite{koch2007charge}. For the parameters of our device, near the nominal lattice location , the change in interaction $U$ is roughly $\delta U \approx +0.75$\,MHz per $+100$\,MHz of change in $\omega_{01}$. Typical $U \sim 2\pi\times -255$\,MHz, except site $Q_4$ which has $U \approx 2\pi\times -160$\,MHz due to a fabrication defect. Despite of this defect, the lattice remains in the strongly interacting regime with $|U|\gg J$. For experiments shown in this work, this defect has little affect on the stabilization performance or the hole dynamics.

The energy spacings of higher transmon qubit levels (i.e. higher occupancy numbers of the lattice site, $n>2$) give rise to effective multibody on-site interactions $U_n$ ($n>2$). For example, at the nominal lattice frequency, the transmon $n=3$ state leads to an effective on-site three-body interaction of $U_3 \approx -27$\,MHz for our qubit parameters. However, these higher order interaction terms are irrelevant for experiments presented in this work where the on-site occupancies are confined to $n=0,1,2$ and probabilities of $n>2$ (e.g. due to far off-resonant excitation, or thermal noise) remain mostly negligible.

\subsection{Lattice spectroscopy}
\label{SI:sub-lattice_spec}

\begin{figure}
    \includegraphics[width=0.50\textwidth]{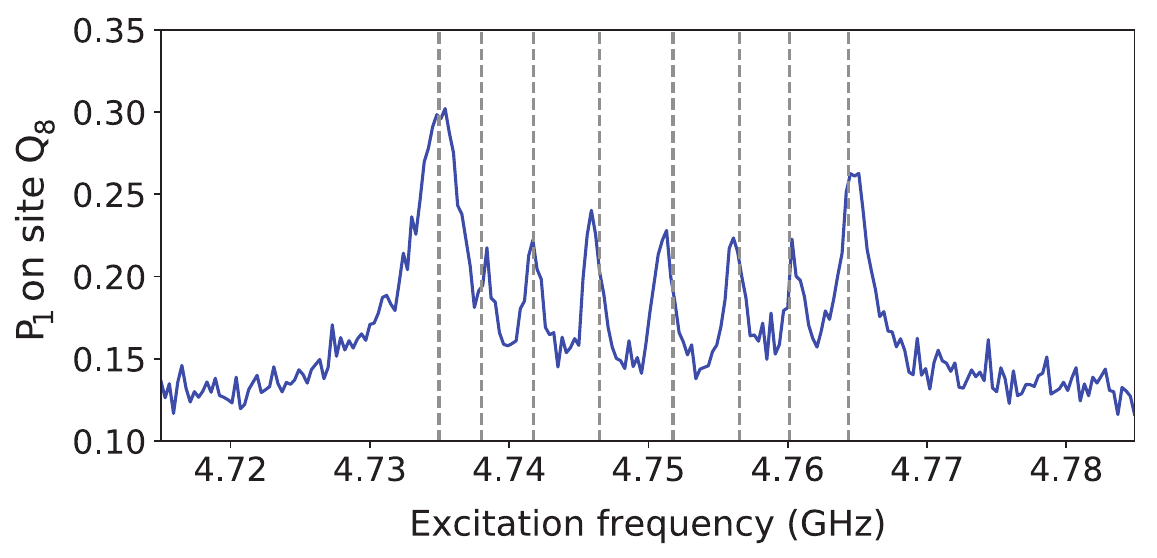}
    \caption{\textbf{Spectroscopy of the lattice.} All 8 lattice sites are tuned to $\approx 4.750$\,GHz. After a weak driving of the lattice via the shared charge driving line, the population in $Q_8$ is measured as a function of excitation frequency. Dotted lines indicate normal mode frequencies calculated from a tight binding model.}
    \label{fig:supp_latt_spec} 
\end{figure}

Our circuit lattice can be easily excited with a coherent microwave tone near the on-site energy, and the site-resolved readout allow spatially resolved spectroscopy of the lattice. As an example, we show in Fig.~\ref{fig:supp_latt_spec} the measured ground band spectra of the 8-site homogeneous lattice ($Q_1 - Q_8$): a weak $2\,\mu$s charge driving pulse is applied to the readout transmission line, simultaneously exciting all sites of the initially empty lattice. We measure the $n=1$ population on $Q_8$ and resolve all 8 eigenmodes of the lattice, with good agreement to predictions from a tight binding model using the measured tunneling rates. Such spectroscopic measurements can be easily extended to transport measurements of the many-body states; and used for Hamiltonian tomography to extract lattice properties, including topological ones, from reflection and transmission spectra~\cite{ma2017hamiltonian}. In our case using the measured mode frequencies and the known tunneling rates, we can calculate the exact on-site disorders and then compensate with the flux biases.

\subsection{Finite thermal occupancies}
\label{SI:sub-n_thermal}

At thermal equilibrium, all resonators in our superconducting circuit remain at finite temperatures. The effective temperatures are typically higher then the actual temperature at base ($20$\,mK) and limited by other sources of microwave noises e.g. leakage of thermal radiation from hotter stages of the fridge. 
We discuss below the effect of thermal population on the stabilization experiments, from (1) the transmon sites, (2) the readout resonators, and (3) the reservoir (lossy resonator).

The thermal population of the transmon qubits at equilibrium are measured in Sec.~\ref{SI:readout+errorbar}, and corresponds to effective temperatures in the range of $\sim 60-80$\,mK. The performance of the dissipative stabilizers are not affected by the thermal population on the stabilized lattice sites, as long as the rate at which the stabilizer is refilling the lattice sites is much faster then the rate at which the on-site thermal relaxation takes place. The latter happens at $\sim n_{\mathrm{th}}^{Q_i}\Gamma_1 < 1$\,kHz, which is much smaller than the optimal stabilizer filling time scales (MHz) for all current experiments. Therefore the dissipative stabilizer could be used to effectively ``cool'' the system to lower entropy per site from the initial thermal ground state.

The primary effect of the readout resonator finite temperature is to induce additional dephasing of the lattice sites due to the dispersive coupling between the resonator and the transmon. In our experiments when the sites are tuned near the lattice location, the large qubit-resonator detuning ($\delta \sim 1.5$\,GHz) and the relatively small coupling ($g \approx 65$\,MHz) make this resonator thermal population induced dephasing rate on the order of a few kHz, much smaller compared to the flux noise induced dephasing which remains as the limiting factor for $T_2^*$.

\begin{figure}
    \includegraphics[width=0.50\textwidth]{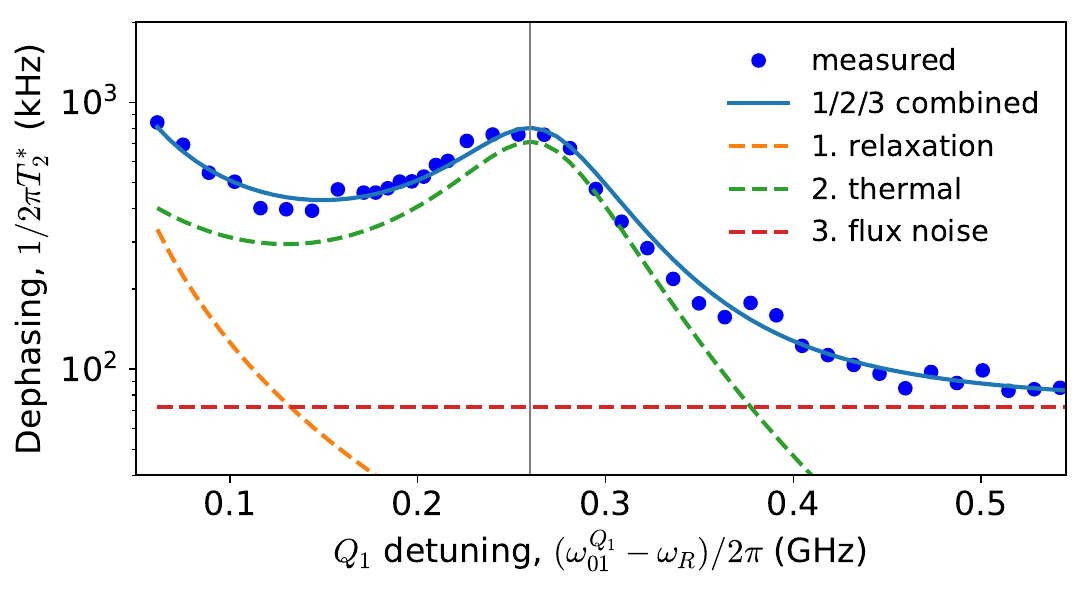}
    \caption{\textbf{Extracting reservoir thermal population.} The dephasing rate of $Q_1$ is measured as a function of detuning from the reservoir. The combined rate (blue) has contributions from qubit relaxation, reservoir thermal population (here calculated with $n_{\mathrm{th}}^R=0.075$), and flux noise on the qubit. The gray vertical line indicates detuning of $U$, i.e.\ when $\omega_{12}^{Q_1}=\omega_R$.}
    \label{fig:supp_lossy-res_nth} 
\end{figure}

The most critical finite temperature effect in our system, is the finite thermal population of the reservoir. The dissipative stabilization relies on the cold reservoir to dissipate the excess entropy. Thermal population in the reservoir will be coupled back into the stabilized sites and lead to infidelities. In Sec.~\ref{SI:numerics-thermalpop} below, we numerically study the effect of reservoir temperature on the stabilization fidelity. The reservoir thermal population $n_{\mathrm{th}}^R$ is calibrated by measuring the additional dephasing induced on $Q_1$ as $Q_1$'s frequency is tuned close to that of the reservoir. In Fig.~\ref{fig:supp_lossy-res_nth}, we plot the measured dephasing rate as a function of the detuning between $Q_1$ and the reservoir. The dephasing has contributions from (1) $T_1$ relaxation, Purcell limited by the reservoir linewidth; (2) fluctuations due to the reservoir thermal population $n_{\mathrm{th}}^R$~\cite{clerk2007thermal}; and (3) flux noise which here we assume be to a constant using measured asymptotic values at larger detunings. The only free parameter for the theory curve is $n_{\mathrm{th}}^R$ which we fit to the experimental data to extract a $n_{\mathrm{th}}^R = 0.075(5)$.

\section{Numerical Simulation of the Stabilizers}
\label{SI:numerics}

We simulate the performance of the single site stabilizer by numerically solving the Lindblad master equation:
\begin{align*}
    \dot{\rho}(t) = - \frac{i}{\hbar}\left[H(t), \rho(t) \right] + \sum_m \frac{1}{2}\left[ 2 C_m \rho(t) C_m^\dagger - \rho(t) C_m^\dagger C_m - C_m^\dagger C_m \rho(t) \right]
\end{align*}
Here the system Hamiltonian $H(t)$ includes both the Bose-Hubbard terms ($H_{\textrm{BH}}$) and the external coherent driving terms for the stabilization, $\rho$ is the density matrix of the lattice system. The collapse operators are $C_m = \sqrt{\gamma_m} L_m$, where $L_m$ are the Lindblad jump operators through which the system couples to the environment and $\gamma_m$ are the corresponding coupling rates. In the simulations we include single photon relaxation $\Gamma_1$ and dephasing $\Gamma_\phi$, reservoir linewidth $\kappa_R$, and finite qubit and reservoir temperatures (with thermal populations $n_{\textrm{th}}^{Q_i}$ and $n_{\textrm{th}}^R)$.

We expect the reservoir loss (from coupling to a 50\,$\Omega$ terminated transmission line) and the qubit $T_1$ relaxation to be Markovian such that the Lindblad master equation applies. The qubit dephasing has contributions from non-Markovian flux noise that is not fully described by a simple Lindblad terms. However for experiments in this work the relatively small dephasing does not affect the physics. For future experiments, We plan to investigate the flux noise spectra for a more complete characterization of the dephasing.

For the numerics shown below, the dissipative stabilizer is modelled after our physical implementation as a multilevel linear oscillator, and thus should capture all relevant experimental features. In our theory paper~\cite{ma2017autonomous}, we developed an effective model for the stabilizer (characterized by an effective refilling rate, and a stabilization infidelity), which could allow effective modelling of more complicated systems as the cost of potentially missing some higher order effects due to approximations made in the effective model.

\subsection{Single site stabilizer performance}
\label{SI:numerics-single_site_stab}

\subsubsection{``One transmon'' scheme}

For the ``one transmon'' scheme, the complete system Hamiltonian used for simulations reads (in the rotating frame of the stabilization drive):
\begin{align*}
    H = (\omega_{R}-\omega_d) a_R^\dagger a_R + (\omega_{01}^S-\omega_d) a_S^\dagger a_S + \frac{U_S}{2} a_S^\dagger a_S^\dagger a_S a_S - J_{1R} (a_S^\dagger a_R + a_R^\dagger a_S) + \frac{\Omega_d}{2} (a_S+a_S^\dagger)
\end{align*}
here the stabilized site $S$ is $Q_1$. The collapse operators included are:
\begin{align*}
    & \textrm{Reservoir dissipation \& thermal population:} & \sqrt{\kappa_R (1+n_{\textrm{th}}^R)} a_S + \sqrt{\kappa_R n_{\textrm{th}}^R} a_S^\dagger \\
    & \textrm{Stabilized site relaxation \& thermal population:} & \sqrt{\Gamma_1^S (1+ n_{\textrm{th}}^S)} a_S + \sqrt{\Gamma_1^S n_{\textrm{th}}^S} a_S^\dagger\\
    & \textrm{Stabilized site dephasing:} & \sqrt{2 \Gamma_\phi^S} a_S^\dagger a_S
\end{align*}
In this scheme the detunings are set to $\omega_{01}^S = \omega_R + U_S$. All other parameters used in the simulations are experimentally measured values. The dephasing term is included to match the measured $T_2^*$ for the $0-1$ transition, but does not necessarily give the right dephasing for the higher levels or an accurate representation of the actual characteristics of the dephasing noise in the device. However, given the small dephasing rate $\Gamma_\phi \ll J,U,\Omega_d$ etc., the stabilizer performance is not affected by the exact details of the dephasing. 
When solving the master equation, we keep on-site occupancies up to $n=2$ for both the stabilized site and the reservoir: during the single site stabilization, higher excited state population ($n>2$) in the strongly interacting site should be suppressed in the absence of resonant driving, while the relatively large linewidth of the reservoir should prevent it from being populated by more than one photon at a time.

\begin{figure}
    \includegraphics[height=0.40\textheight]{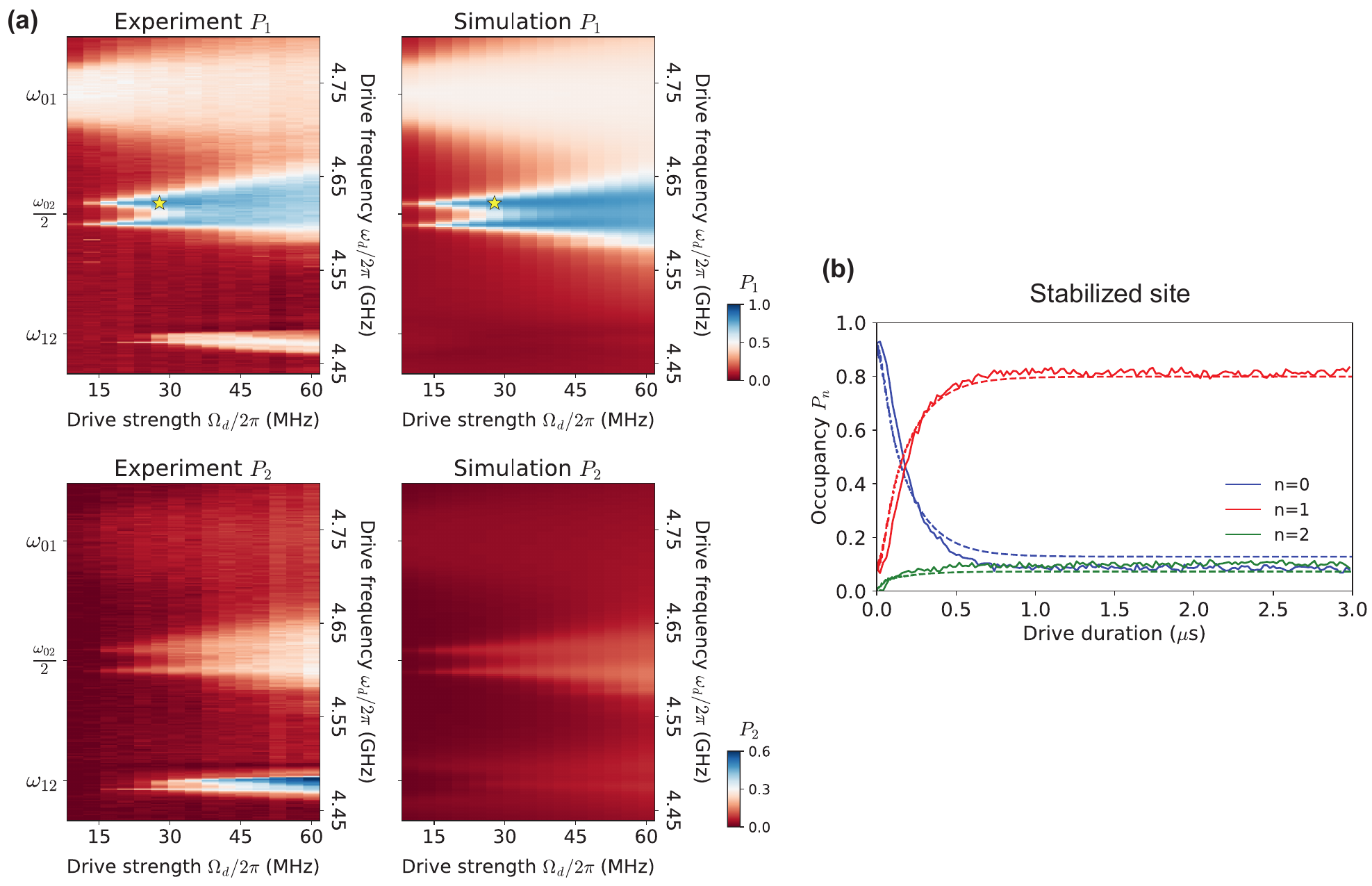}
    \caption{\textbf{``One transmon'' scheme.} Comparison between experimental data and numerical simulations. \textbf{(a)} Stabilized site occupancy after $3\,\mu$s drive. \textbf{(b)} Time dynamics at optimal driving.
    }
    \label{fig:supp_1qScheme-compare} 
\end{figure}

In Fig.~\ref{fig:supp_1qScheme-compare} we plot the stabilized site population using the ``one transmon'' scheme after driving for $3\,\mu$s. We also plot the filling dynamics on the stabilized site at the driving parameters that give the experimentally observed optimal stabilizer performance. We see quantitative agreement between experiment and the no free parameter numerical simulation, with two main discrepancies: (1) At higher driving rates, we measure more $n=2$ population experimentally, which can be attributed to off-resonant excitation of higher transmon levels on the stabilized site that are not included in the numerics. (2) When driving near $\omega_R$, we observe significant population in the stabilized site. This comes from a direct coupling between the stabilization drive line and the reservoir, due to their close proximity on the device chip and the resulting capacitive coupling. The resonant drive populates the reservoir and those photons can tunnel back into the stabilized site. We can qualitatively reproduce the observed features by adding to the simulation a term $\propto \Omega_d \cos{[i(\omega_R-\omega_d)t]} (a_R+a_R^\dagger)/2$, with a strength consistent from finite element simulations of the device. The exact strength and frequency dependence of this coupling is difficult to predict or measure precisely. We have verified numerically that the addition of higher transmon levels, the direct driving term on the reservoir, or the exact implementation of transmon dephasing, all have negligible affects on the stabilizer performance near the designed optimal driving parameters.

\begin{figure}
    \includegraphics[height=0.40\textheight]{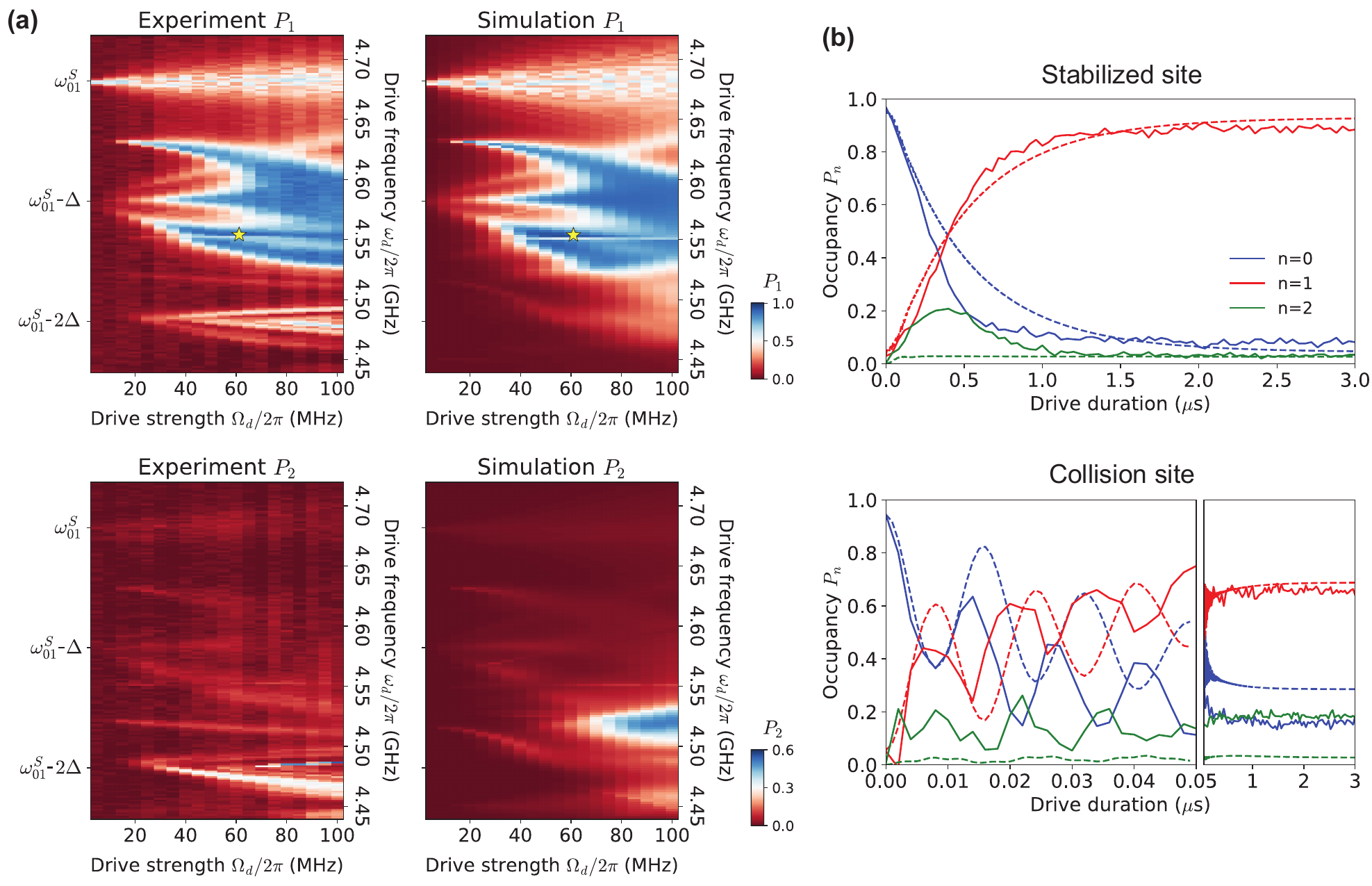}
    \caption{\textbf{``Two transmon'' scheme.} \textbf{(a)} Stabilized site occupancy after $3\,\mu$s drive. \textbf{(b)} Time dynamics of both the stabilized site and the collision site at optimal driving.
    }
    \label{fig:supp_2qScheme-compare} 
\end{figure}

\subsubsection{``Two transmon'' scheme}

For the ``two transmon'' scheme, the Hamiltonian reads:
\begin{align*}
    H = (\omega_{R}-\omega_d) a_R^\dagger a_R + (\omega_{01}^S-\omega_d) a_S^\dagger a_S + (\omega_{01}^C-\omega_d) a_C^\dagger a_C + \frac{U_S}{2} a_S^\dagger a_S^\dagger a_S a_S + \frac{U_C}{2} a_C^\dagger a_C^\dagger a_C a_C\\
     - J_{1R} (a_S^\dagger a_R + a_R^\dagger a_S) - J_{12} (a_S^\dagger a_C + a_C^\dagger a_S) + \frac{\Omega_d}{2} (a_C+a_C^\dagger)
\end{align*}
Here the collision site ($C$) is $Q_1$, the stabilized site ($S$) is $Q_2$, and they are detuned by $\Delta = \omega_{01}^S - \omega_{01}^C = \omega_{01}^C -\omega_R = 100$\,MHz. The collapse operators are the same as for the ``one transmon'' scheme, with the addition of terms for the collision site that have the same expressions as those for the stabilized site. The stabilization drive line, while driving $Q_1$($C$) at rate $\Omega_d$, also has a direct coupling to $Q_2$($S$) with rate $\eta \Omega_d$ due to residual capacitive coupling on chip. Therefore to accurately simulate the stabilized site dynamics near $\omega_d \sim \omega_{01}^S$, we include in $H$ an additional term $\eta \Omega_d \cos{[i(\omega_{01}^S-\omega_d)t]} (a_S + a_S^\dagger)/2$. We use $\eta = -0.103$ measured experimentally from resonant Rabi rates of $Q_2$. The inclusion of this term does not effect the optimal stabilizer performance, where the drive is always off-resonant with $\omega_{01}^S$.

The comparison between experiment and simulation for the ``two transmon'' scheme is shown in Fig.~\ref{fig:supp_2qScheme-compare}. The direct resonant driving at $\omega_d = \omega_{01}^C = 2\pi\times 4.683$\,GHz is still visible with signs of coherent oscillations around $P_1 \sim 0.5$. The expected stabilization peak at the collision site frequency $\omega_d = \omega_{01}^C = \omega_{01}^S-\Delta$ is accompanied by other features with high $P_1$ from higher-order collision processes~\cite{ma2011photon}:

\medskip \textbf{\noindent Higher order stabilization processes.}
The peak appearing near $(\omega_{01}^S-\Delta/2)$ corresponds to a 2-photon process that populates the stabilized site and the collision site, the latter now serving as the lossy channel due to its coupling to the lossy resonator site. The feature close to $(\omega_{01}^S-3\Delta/2)$ arises from fourth-order population of the stabilized site in $n=1$ with $4\times \omega_d = \omega_{01}^S + 3\times\omega_R$. Both features experience substantial drive-power dependent Stark shifts. The measured optimal single site stabilizer fidelity is $P_1 = 0.89(+0.04/\text{\textminus}0.01)$ at $\Omega_d = 2\pi\times 60$\,MHz, $\omega_d = 2\pi\times 4.555$\,GHz. On top of this broad stabilization feature, a sharp dip in fidelity occurs at $\omega_d = \omega_{01}^S+U/2$ resulting from quantum interference between the intended process and the direct ``one transmon''  process on the stabilized site. Both the measured steady-state fidelity and the stabilizer dynamics are in quantitative agreement with numerical simulation, with the highest observed fidelity primarily limited by reservoir thermal population.

The observed population near $\omega_d \sim \omega_R$ is again from the direct driving of the reservoir, not included in the simulations. There is also a discrepancy at high $\Omega_d$ near $\omega_d \sim \omega_{01}^S - 3\Delta/2$ which is likely a result of state truncation in the simulation: the higher excited states of the collision site lie around this frequency, and can get populated at high driving rates. We also show in \textbf{(b)} the filling dynamics at the optimal driving parameter (star in \textbf{(a)}). At short times, the collision site population show oscillations at rate $\sim \Omega_d$ from direct driving by the stabilization drive. Since the optimal point happens near a sharp feature resulting from interference of multiple processes, the oscillatory dynamics are sensitive to exact on-site energies and various driving rates, resulting in some discrepancy between data and simulation. The excess $n=2$ population in the measured steady state could be a result of off-resonant driving to higher $n$ states that ended up (e.g. via decay) in $n=2$. 

\begin{figure}
    \includegraphics[width=0.45\textwidth]{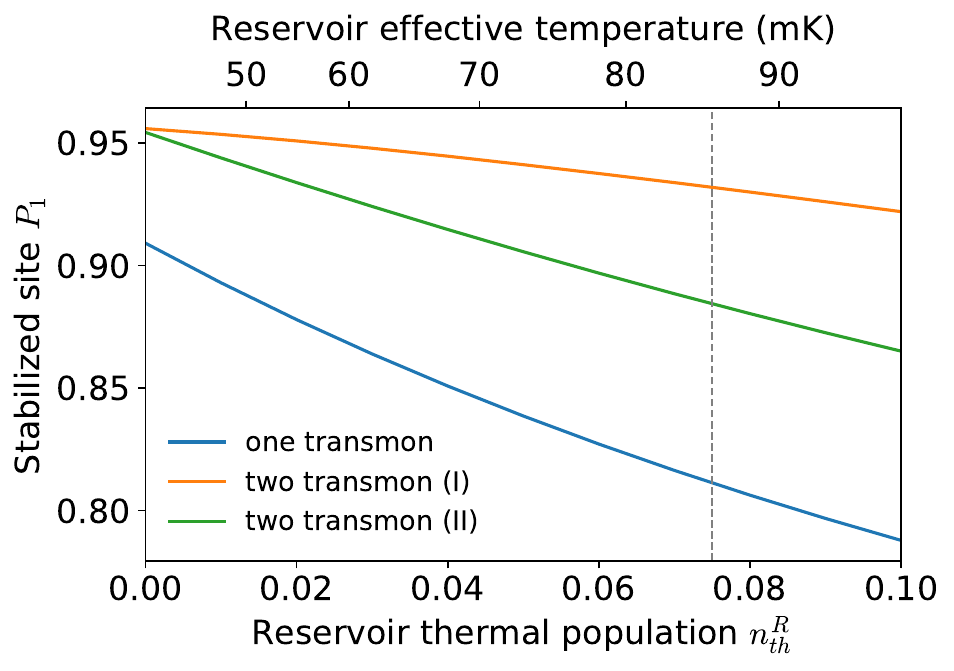}
    \caption{\textbf{Effect of reservoir temperature.} As a function of thermal population in the reservoir, the optimal single site stabilization fidelity is calculated using the current device parameters, for both the ``one transmon'' and ``two transmon'' scheme. For the ``two transmon'' scheme the detuning is set to $\Delta=100$\,MHz, and we show optimal fidelities both when driving near $\omega_d\sim \omega_{01}^C$ (II), and driving at the experimentally observed optimal $\omega_d$ near the sharp feature at slightly below $\omega_d \sim \omega_{01}^S + U/2$ (I). The gray line indicates $n_{\mathrm{th}}^R = 0.075$ in our current experiments.
    }
    \label{fig:supp_effect-of-nth} 
\end{figure}

\subsection{Reservoir thermal population}
\label{SI:numerics-thermalpop}

The current single site stabilization fidelity (and therefore also the Mott fidelity) is mostly limited by thermal populations in the reservoir that re-enter the stabilized site(s). In Fig.~\ref{fig:supp_effect-of-nth}, we plot the simulated optimal fidelities for the ``one transmon'' and ``two transmon'' schemes as a function of the reservoir thermal population $n_{\mathrm{th}}^R$. As $n_{\mathrm{th}}^R$ is varied, the optimal driving parameters remain unchanged. Qualitatively, the numerical results indicate a thermal population of $n_{\mathrm{th}}^R$ in the reservoir causes at least the same amount of additional infidelity in the stabilizer compared to a reservoir at near zero $n_{\mathrm{th}}^R$.

In our current device and microwave setup, the reservoir at $\sim 4.5$ GHz with measured $n_{\mathrm{th}}^R = 0.075$ corresponds to an effective temperature of $\approx 80$\,mK, significantly higher than the physical temperature of the sample at $\sim 20$\,mK. We expect the reservoir temperature to be limited by thermal photon from the upper stages of the fridges, since the reservoir's linewidth is obtained by coupling to the input-outline transmission line of the sample. Additional filtering/attenuation on the input and output lines at the reservoir frequency should help reduce $n_{\mathrm{th}}^R$: at $\sim 45$\,mK, $n_{\mathrm{th}}^R$ already drops to $1\%$ at $\omega_R\sim 4.5$\,GHz. In the future, the reservoir thermal population can be further reduced by moving the reservoir to higher frequencies, and by coupling the reservoir to an individual terminated line that is well thermalized to the base of the fridge.

\begin{figure}
    \includegraphics[width=0.55\textwidth]{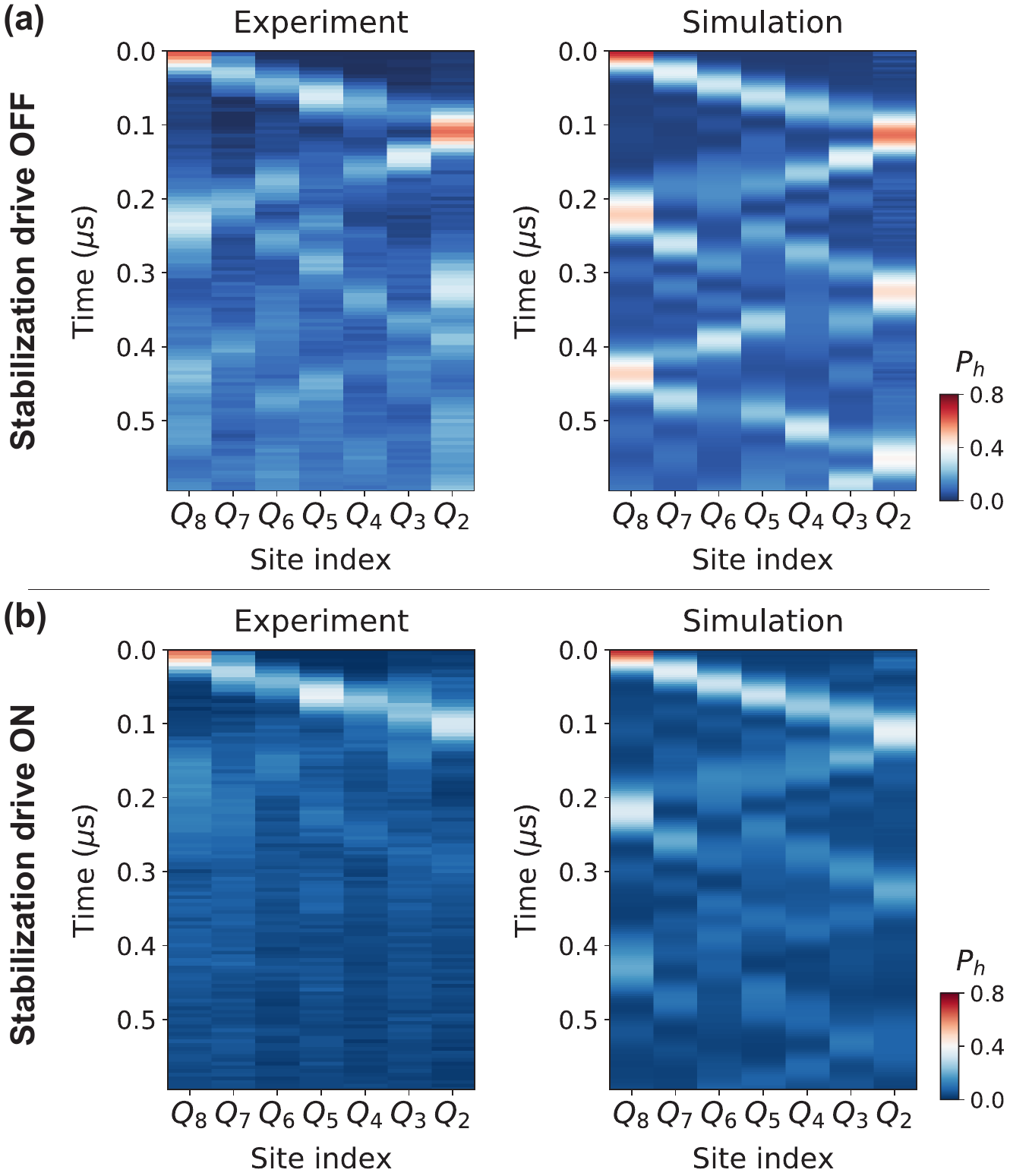}
    \caption{\textbf{Dynamics of hole defect.} Comparison between experimental data and numerical simulations, for evolution of a single hole defect on top of the Mott insulator, in the absence (a) or presence (b) of the stabilization drive. 
    }
    \label{fig:supp_hole-dyn} 
\end{figure}

\subsection{Hole dynamics}
\label{SI:numerics-hole_dynamics}

Here we simulate the dynamics of the hole defect on top of the dissipatively prepared Mott insulator (Fig.~5 in main text). The stabilizer used is the ``two transmon'' scheme above, and the coherent lattice is added to the simulation assuming degenerate on-site energy and using experimentally measured tunnelings $J_{ij}$. The simulated hole dynamics with and without the stabilization drive are plotted and compared to the experimental data in Fig.~\ref{fig:supp_hole-dyn}. The initial state used in the simulation has unity filling ($P_1=1$) across the lattice expect on $Q_0$ which is empty ($P_0=1$). The additional ``dephasing'' of the wavepacket in the experiments is partly a result of the initial infidelity of the dissipatively prepared Mott state (thermal mixture of $P_0\approx 10$\% on each site). The initial hole density of $\sim 80$\% is a result of the relatively small detuning ($\sim 25$\,MHz) between $Q_8$ and the rest of the lattice during the Mott preparation, and for ease of comparison we have scaled the simulations to the same hole density.

\end{document}